\documentclass[12pt, twoside, here]{article}
\usepackage{epsf}
\usepackage{times,colordvi,amsmath,epsfig,float,color,multicol}
\usepackage{graphics}
\usepackage{hhline}
\usepackage[large]{subfigure}
\usepackage[latin1]{inputenc}
\usepackage{rotating}
\usepackage{marvosym}
\usepackage{amsfonts}
\usepackage{amssymb}

\title{Theoretical Levels of Control as a Function of Mean Temperature and Spray Efficacy in the Aerial Spraying of Tsetse Fly}

\renewcommand{\thefootnote}{\fnsymbol{footnote}}

\author{S. J. Childs \\ \\ {\small\em ARC-Onderstepoort Veterinary Institute, Private Bag X5,} \\ {\small\em Pretoria, 0110, South Africa.} \\ {\small\em Tel: +27 72 8459556 \ \ Email: simonjohnchilds@gmail.com}}

\date{Acta Tropica, 117: 171--182, 2011}

\begin{document}

\maketitle
\renewcommand{\thefootnote}{\arabic{footnote}}

\begin{abstract}
\noindent {\em The hypothetical impact of aerial spraying on tsetse fly
populations is investigated. Spray cycles are scheduled at intervals two days
short of the first interlarval period and halted once the last of the female
flies that originated from pre-spray-deposited pupae have been sprayed twice.
The effect of temperature on the aerial spraying of tsetse, through its
reproductive cycle and general population dynamics, is of particular interest,
given that cooler weather is preferred for the settling of insecticidal
droplets. Spray efficacy is found to come at a price due to the greater number
of cycles necessitated by cooler weather. The extra cost is argued to be worth
while. Pupae, still in the ground at the end of spraying, are identified as the
main threat to a successful operation. They are slightly more vulnerable at the
low temperature extreme of tsetse habitat (16 $^\circ$C), when the cumulative,
natural pupal mortality is high. One can otherwise base one's expectations on
the closeness with which the time to the third last spray approaches one
puparial duration. A disparity of anything close to the length of a spray cycle
advocates caution, whereas one which comes close to vanishing should be
interpreted as being auspicious. Three such key temperatures, just below which
one can anticipate an improved outcome and just above which caution should be
exercised, are 17.146 $^\circ$C, 19.278 $^\circ$C and 23.645 $^\circ$C. A
refinement of the existing formulae for the puparial duration and the first
interlarval period might be prudent in the South African context of a sympatric
Glossina brevipalpis-G. austeni, tsetse population. The resulting aerial
spraying strategy would then be formulated using a G. brevipalpis puparial
duration and a G. austeni first interlarval period.} 
\end{abstract}

Keywords: Tsetse; {\em Glossina}; aerial spraying; insecticide; deltamethrin; trypanosomiasis; sleeping sickness. 

\section{Introduction}

Aerial spraying has been in use for more than 50 years as a means of controlling
and even eradicating tsetse fly ({\em Glossina}: Diptera), according to
Allsopp\nocite{Allsopp1} (1984). One of the earliest and most notable triumphs
was the complete eradication of {\em G. pallidipes} from vast habitat in
KwaZulu-Natal, in the middle of the twentieth century. Although a whole
spectrum of counter measures was resorted to in that campaign (\mbox{Du
Toit}\nocite{DuToit}, 1954), it was ultimately reliant on aerial spraying.
Unfortunately, \mbox{Du Toit}\nocite{DuToit} (1954) also directly attributes
much of its success to the use of D.D.T. in the form of a thermal aerosol, or
smoke. This he described as being semi-gaseous, sinking through the thick
forest canopy and into dense undergrowth, above the ground, ``within seconds''.
The modern operation conventionally utilizes a relatively harmless pyrethroid
such as endosulfan or deltamethrin (Allsopp\nocite{Allsopp1}, 1984). An aerosol
of insecticide is discharged from a formation of aircraft, flying at low
altitude (less than 100 metres a.g.l.) and guided by G.P.S. The concentration of
the active ingredient introduced into the environment in this way is, typically,
very small, with the result that there is little residual effect, even one day
after its application. Adult flies are extremely susceptible to the insecticide
and kill rates close to 100\% can be anticipated under favourable conditions.
The main challenge to controlling tsetse by way of aerial spraying, is that the
pupal stage is spent underground, where it is largely protected from
insecticides. Repeat cycles therefore need to be scheduled in order to kill the
new flies which begin emerging the day after spraying. Both economic and
environmental considerations obviously dictate that the number of such cycles be
minimised, however, the problem is that if the time between spray cycles is too
long, recently ecloded flies will themselves become reproductive and deposit
pupae. 

The reproductive life cycle of the fly makes it especially well disposed to
control through the cyclical application of insecticides since all developmental
periods are entirely temperature dependent, including the regularity with which
a single pupa is deposited and develops. The pupal duration and the time between
female eclosion and the deposition of her first pupa are therefore readily
predictable, given that the temperature is known. It is in this fact that there
lies a major vulnerability open to exploitation for the control of tsetse. A
repeat spray cycle can be scheduled just prior to when the more recently ecloded
flies, themselves begin to deposit pupae. The cycles are halted only once the
last of the flies to emerge from pre-spray-deposited pupae have been sprayed,
ideally, twice. At 25 $^\circ$C, for example, the time from eclosion to the
deposition of the first larva takes 16 days. One would therefore spray at 14 day
intervals, were one to choose to spray at 25 $^\circ$C. The total time from
parturition to eclosion is 26 days for female pupae. Three spraying cycles
should therefore ensure that all flies, that existed as pupae in the ground
prior to the start of spraying, will have emerged and been subjected to at least
one spray. In practice, a fourth cycle would probably be employed to allow for
the fact that there is significant variance about the mean values of the
puparial duration (Phelps and Burrows\nocite{phelpsAndBurrows1}, 1969). Other
aspects of tsetse population dynamics are also largely temperature-dependent
(\mbox{Hargrove\nocite{Hargrove3}, 2004}), although soil-humidity can play an
as, or more, important role in early mortality, depending on the species
(Childs\nocite{Childs2}, 2009). 

In the event that each spray does indeed kill every fly, the aforementioned
strategy will obviously lead to the eradication of any closed tsetse population.
If, however, the kill rate is very low (e.g. less than 90 \%), substantial
numbers of female flies will survive a spray cycle to deposit pupae before the
next cycle. Such an eventuality will severely jeopardise the success of the
operation and the ultimate goal of eradication will not be achieved. A more
interesting scenario arises when the kill rate is close to, but less than, 100
\%. Under such circumstances it may be that the absolute numbers and the density
of the surviving population are so small that the population will die out
anyway, due to chance effects coupled with the very small probability of a
female meeting and being inseminated by a male (Hargrove\nocite{Hargrove10},
2005). Little theoretical analysis of such marginal situations has been carried
out, despite a long history of aerial spraying for the control of tsetse.
Similarly, there has been little published analysis on how the outcome of an
aerial spraying operation depends on the number of cycles, ambient temperature,
an individual fly's probability of surviving a spraying event, as well as any
interaction between these variables.

The effect of temperature on the aerial spraying of tsetse, through its
reproductive cycle and general population dynamics, is investigated in this
work, given that cooler weather is preferred from a point of view of spray
efficacy. The effects of temperature on spray efficacy are, however, not
modelled. Spray efficacy is, instead, the context of this investigation. The
dynamics of a population during an aerial spraying campaign are crudely modelled
using published information on the temperature dependent rates of mortality, for
different stages of the life cycle, and of development and pupal production. The
hypothetical impact of aerial spraying on a tsetse fly population is determined.
The idea is to establish some of the conditions which will be sufficient for a
successful outcome in the context of a closed population (re-invasion being an
ever present threat which will ultimately compromise even the most successful
aerial spraying operation). The study undertaken is an hypothetical one
pertaining to some, general, geographic region, of approximately the size,
population densities and temperatures one might expect. No accompanying degree
of precision can, accordingly, be given. One would expect such precision to be
profoundly dependent on the variance in mean daily temperature pertaining to a
specific region. Optional stochastic features in the model would only be
employed in a specific case study, one in which the variance in temperature and
other parameters for the environment in question, were known. While a stochastic
treatment would allow risk to be determined, the computational
overheads are also, obviously much higher. For the present, this work is
exploratory. 

Although the work is widely applicable, the South African scenario of a
sympatric, {\em Glossina brevipalpis}-{\em G. austeni} population is of
particular and immediate interest. This study has been brought about by a recent
revival of interest in tsetse control in South Africa, since the eradication of
{\em G. pallidipes} in the times of \mbox{Du Toit\nocite{DuToit}} (1954). It is
not known what the effect of rugged terrain, the associated winds and the level
of modern insecticide penetration into forested, riverine habitats will be in
the case of these species and so this study forms just one, small part of a
recently commissioned, suite of work relating to various avenues of tsetse
control. That work could be said to have commenced with Kappmeier
Green\nocite{KappmeierGreen} (2002) and it includes a comprehensive
investigation of the use of odour-baited targets and pour-ons e.g.
Childs\nocite{Childs3} (2010) and Esterhuizen et
al.\nocite{EsterhuizenKappmeierGreenNevillVanDenBossche} (2006), to mention only
two. It also includes studies of vector competence, such as Motloang et
al.\nocite{MotloangMasumuVanDenBosscheMajiwaLatif} (2009), as well as studies of
habitat e.g. Childs\nocite{Childs2} (2009), Esterhuizen et
al.\nocite{EsterhuizenKappmeierGreenMarcottyVanDenBossche} (2005) and
Hendrickx\nocite{Hendrickx} (2007). The possibility of employing the sterile
insect technique in the South African context has also been mooted by way of
Kappmeier Green et al.\nocite{KappmeierGreenPotgieterVreysen} (2007). Some work
even goes so far as to raise the possibility of competition from tsetse species
which lack the same vector competence, while Esterhuizen and van den
Bossche\nocite{EsterhuizenVanDenBossche} (2006) entertain the possibility of
protective netting. It is in this context that the present study on aerial
spraying should be seen.

In what follows, a brief overview of the reproductive life cycle of
the tsetse fly is given and the strategy for the aerial application of
insecticide is outlined. Although the resulting model itself is capable of
accepting variable temperature inputs, results are presented in terms of
constant temperature; for obvious reasons. These results are analysed and a
constant-temperature formula, which is a good approximation of the outcome, is
derived. 

\section{The Temperature Dependent Life-Cycle of the Tsetse Fly} \label{lifeCycle}

Pupae are deposited in the ground where they remain for a period of time. This
period, the period between larviposition and the emergence of the first imago,
is known as the puparial duration, $\tau_0$. The puparial duration is a function
of temperature, $T$. Pupae also die off at some temperature-dependent, daily
rate, $\delta_0$, and those flies which subsequently emerge have a probability
$\gamma$ of being female. During the first few hours, the young, teneral fly's
exoskeleton is soft and pliable, its fluid and fat reserves are at their lowest
and a first blood meal is imperative for its survival. It is at this time that
the insect is at its most vulnerable and it is also at this time that its
behaviour is least risk averse (Vale\nocite{Vale3}, 1974). post-pupal survival
can be defined as $e^{- \delta_1}$ per day at the pre-ovulatory stage (the time
between female eclosion and ovulation). Thereafter the female tsetse fly's
chances of survival are higher and can be defined as $e^{ - \delta_2}$ per day.
The female tsetse fly mates only once in her life with the chance $\eta$ that
she is successfully inseminated (in the current context, $\eta$ is taken as
unity unless stated otherwise). The time between female eclosion and the
production of the first pupa is known as the first interlarval period, $\tau_1$.
Thereafter she produces pupae at a shorter interlarval period, $\tau_2$. 

\subsection{The Puparial Duration}

In the moments immediately following parturition, the third-instar larva
burrows into the substrate (sand or leaf litter) and secretes a puparial case.
It remains there for a period of time known as the puparial duration, during
which it goes through all the developmental changes (including the sensu strictu
pupal stage) required to produce the adult. The puparial duration is
temperature-dependent and may be predicted using the formula \begin{eqnarray}
\label{31} \tau_0 &=& \frac{ 1 + e^{a + bT} }{k}  \end{eqnarray}  (Phelps and
Burrows\nocite{phelpsAndBurrows1}, 1969). For females, $k = 0.057 \pm 0.001$, $a
= 5.5 \pm 0.2$ and $b = -0.25 \pm 0.01$ (Hargrove\nocite{Hargrove3}, 2004).
These coefficients are considered preferable to the original ones as the
original fit of Phelps and Burrows\nocite{phelpsAndBurrows1} (1969) made use of
data for temperatures well below \mbox{$16 \ ^\circ\mathrm{C}$}, temperatures at
which pupal fatalities are exceptionally high. 

{\sc Remark:} Notice that for temperatures below $22 \ ^\circ\mathrm{C}$ the
exponential term is significant and the puparial duration begins to lengthen
dramatically (Table \ref{durations}). 

Parker\nocite{Parker1} (2008) reports that the puparial durations of all
species, with the exception of {\em G. brevipalpis}, are thought to lie within
10\% of the value predicted by this formula. {\em G. brevipalpis} takes a little
longer. For the same conditions which produce a {\em G. morsitans} puparial
duration of 30 days, {\em G. brevipalpis} has a puparial duration of 35 days.
This has important implications for the aerial spraying of {\em G. brevipalpis}.
The shortest puparial duration is that of {\em G. austeni}. {\em G. austeni}'s
puparial duration was 28 days under the aforementioned conditions. These
observations are noteworthy given the South African context of a sympatric, {\em
G. brevipalpis}-{\em G. austeni} population.

\begin{table}[H]
\begin{center}
\begin{tabular}{c|c c c c c c c c}  
&  &  &  &  &  &  &  &  \\
T & \ $16 \ ^\circ\mathrm{C}$ & $18 \ ^\circ\mathrm{C}$ & $20 \ ^\circ\mathrm{C}$ & $22 \ ^\circ\mathrm{C}$ & $24 \ ^\circ\mathrm{C}$ & $26 \ ^\circ\mathrm{C}$ & $28 \ ^\circ\mathrm{C}$ & $30 \ ^\circ\mathrm{C}$ \\ \\ \hline \\
$\tau_0$ / $\mathrm{days}$ & 96 & 65 & 46 & 35 & 28 & 24 & 21 & 20 \\ \\
$\tau_1$ / $\mathrm{days}$ & 22 & 20 & 19 & 18 & 16 & 15 & 14 & 14 \\ \\
$\tau_2$ / $\mathrm{days}$ & 16 & 14 & 12 & 11 & 10 & 9 & 8 & 7 \\ \\
\end{tabular}
\caption{The puparial duration, $\tau_0$, the first interlarval period, $\tau_1$, and second interlarval period, $\tau_2$, at different temperatures (calculated according to Hargrove\nocite{Hargrove3}, 2004).} \label{durations}
\end{center}
\end{table}

\subsection{The First and Subsequent Interlarval Periods}

Female tsetse produce only one larva at a time. This larva can weigh more than
the fly that deposits it, implying that the female must make a considerable
investment in the larva and that the reproductive rate is very much lower than
in most insects. The teneral female which emerges from the puparium also lacks
the fully developed flight musculature of the mature fly, and has low fat
reserves. Flight muscle and fat need to be built up to mature levels before
larval production can begin and the time between female eclosion and the
production of the first pupa, $\tau_1$, is accordingly longer than the
subsequent interlarval periods, $\tau_2$. The female must also be inseminated
before larval production can start. The female generally mates only once in her
life and this normally happens in the first week of adult life before flight
muscle development is complete (Hargrove\nocite{Hargrove3}, 2004). The effect of
temperature on both periods has only been estimated once in the field and then
only for {\em G. pallidipes}. The predicted mean time taken from female eclosion
to the production of the first pupa is obtained using $k_1 = 0.061 \pm 0.002$
and $k_2 = 0.0020 \pm 0.0009$ (Hargrove\nocite{Hargrove4}, 1994 and
1995\nocite{Hargrove5}) in Jackson's formula
\begin{eqnarray} \label{interlarvalPeriodFormula} 
\tau_i &=& \frac{ 1 }{ k_1 + k_2 \left( T - 24 \right) } \hspace{10mm} i = 1,2 \end{eqnarray} 
(Anonymous\nocite{Anonymous}, 1955). The subsequent interlarval periods are
predicted using $k_1 = 0.1046 \pm 0.0004$ and $k_2 = 0.0052 \pm 0.0001$
(Hargrove\nocite{Hargrove4}, 1994 and 1995\nocite{Hargrove5}). These
coefficients are considered preferable to the original ones as, once below 18
$^\circ\mathrm{C}$, those of Jackson (Anonymous\nocite{Anonymous}, 1955) predict
a second interlarval period, slightly longer than the first. The values obtained
by evaluating the above formulae, for a range of temperatures, are presented in
Table \ref{durations}.

One might surmise that both periods could be shorter than the formula predicts
for {\em G. austeni}, based on the small size of the fly and in keeping with its
shorter puparial duration. A shorter first interlarval period is obviously a
concern for the aerial spraying of {\em G. austeni}. For {\em G. brevipalpis}
one suspects longer periods based on diammetrically opposite arguments. In this
case, however, the only relevance to aerial spraying is economic. 
%Hargrove\nocite{Hargrove3}, 2004's improved {\sc East African
%High Commision} \cite{Anonymous} formulae.

\subsection{Natural Mortality}

Adult tsetse mortality has been found to increase with temperature in both {\em
G. morsitans} and {\em G. pallidipes} (Hargrove\nocite{Hargrove3}, 2004). It has
also been shown to be a function of age in {\em G. morsitans} (Hargrove
\nocite{Hargrove1}, 1990). There is an initial, rapid decrease in mortality once
the newly-ecloded fly has taken one or more blood meals, the flight musculature
has developed and fat reserves have been increased. Pupal mortality is not as
straight forward. Although the effects of both temperature and humidity on pupal
mortality are known to be important, they vary profoundly according to the exact
stage of development and are cumulative, rather than instantaneous
(Childs\nocite{Childs2}, 2009). One might therefore surmise that the age
dependence which characterises post-pupal mortality (observed by
Hargrove\nocite{Hargrove1}, 1990 and 1993\nocite{Hargrove2}) is largely a
consequence of pupal history. \mbox{Du Toit}\nocite{DuToit} (1954) reported
exceptionally high levels of parasitized pupae for both {\em G. brevipalpis} and
{\em G. pallidipes} in the Hluhluwe-iMfolozi Game Reserve (54.8\% and 51.9\%
respectively). This he identified as being almost exclusively the work of the
tsetse-specific bombyliid, {\em Thyridanthrax brevifacies}. He concluded the
parasitism to be density dependent. Chorley\nocite{Chorley} (1929) observed
similar levels of parasitism among G. morsitans pupae in Zimbabwe (reported in
\mbox{Du Toit}\nocite{DuToit}, 1954). The mortality rates of pupae have only
been directly estimated once in the field (Rogers and
Randolph\nocite{RogersAndRandolph1}, 1990). Mortality was largely due to
predation in that instance, particularly by ants. This predation was also found
to be density dependent.  

The definitive theoretical work relating the mortalities of the different stages is that of Williams\nocite{Williams1} et al. 1990. In their equation,
\begin{eqnarray} \label{williamsEquation}
B e^{ - \tau_{0}(\delta_{0} + R) - \tau_{1}(\delta_{1} + R) -
\tau_{2}(\delta_{2} + R)} &=& 1 - e^{ - \tau_{2}(\delta_{2}
+ R)},
\end{eqnarray}
$R$ is the growth rate and $B$ is the fecundity. It was decided to model fly
mortality by assigning the pre-ovulatory-stage cohorts mortality rates twice
the adult values. Rogers and Randolph\nocite{RogersAndRandolph1} (1990) and
Childs\nocite{Childs2} (2009) were ultimately used as a guide for a minimum
(worst-case) value for pupal mortality, although it is recognised that there
are causes of pupal mortality other than predation and water loss, including fat
loss and parasitism. Whereas two of the aforementioned are undoubtedly the
effects of temperature and humidity, after much debate it was decided to ignore
the dependence for the present purposes. Over-complication and a loss of
generality to all species were the major considerations motivating this
decision. A constant value as conservative as 0.01 $\mathrm{day}^{-1}$
(Hargrove\nocite{Hargrove11}, 2009) was therefore used for pupal mortality at
all temperatures. It is also low in light of the average pupal mortality
\mbox{Du Toit}\nocite{DuToit} (1954) recorded for {\em G. pallidipes} (62.5\%)
and that observed by Chorley\nocite{Chorley} (1929) for {\em G. morsitans} (60\%
to 40\%). Whereas the value of 0.01 $\mathrm{day}^{-1}$ may underestimate
immature losses, particularly in hot, dry weather, it produces the desired
result of a worst-case scenario for the success of an aerial spraying operation.
Solving Equation \ref{williamsEquation}, using Newton's method and assuming a
pre-existing equilibrium involving the aforementioned parameters, suggested the
Table \ref{parametersUsed} values. A miscarriage rate of 5\%, and therefore a
fecundity of 0.475 was used (in keeping with Williams et al.\nocite{Williams1}
1990). 

Although the fly mortalitities so generated represent equilibrium values in
terms of \mbox{Equation \ref{williamsEquation}}, they constitute a worst-case,
aerial spraying scenario in terms of the values obtained by Hargrove and
Williams\nocite{HargroveAndWilliams}, 1998 (Hargrove\nocite{Hargrove11}, 2009).
The assumption of a worst-case scenario in the pupal instance is far more
important. The pupa is not subject to the high insecticidal mortalities which
tend to render any subsequent natural mortality irrelevant. A conservative pupal
mortality leads to high rates of eclosion and compensates, to an extent, for the
omission of any density dependence from the model. Any density dependence might
be expected to manifest itself fairly uniformly across the temperature range. At
worst, the lack of it might slightly magnify the differences between the final
results. Variables such as vegetation index and soil humidity are also regarded
to vary (and therefore be relevant) only in the medium to long term. Some
comfort can be taken from the knowledge that the effects of natural mortalities
are all very small in comparison to those due to aerial spraying. They have
little bearing on the overriding trends reported in this work and to a certain
extent, this knowledge permits the primitive approach taken. 

The accumulated mortality described above can be modelled linearly as 
\begin{eqnarray} \label{26} 
\delta & = & \left\{ 
\begin{array}{l} \delta_0 \ t  \\ 
\delta_1 \left[t  - \tau_0\right] + \delta_0 \tau_0  \\ 
\delta_2 \left[ t - (\tau_1 - \tau_2) - \tau_0 \right] + \delta_1 \left[ \tau_1 - \tau_2 \right] + \delta_0 \tau_0 
\end{array} \hspace{5mm} \mbox{for} \hspace{5mm} 
\begin{array}{rcl}
&t&< \ \tau_0 \\
\tau_0 \ \le&t&< \ \tau_1 - \tau_2 + \tau_0 \\ 
&t&\ge \ \tau_1 - \tau_2 + \tau_0, 
\end{array} 
\right. \nonumber \\
\end{eqnarray} 
where $t$ denotes age, for the present. The model parameters at different
temperatures are presented in Table \ref{parametersUsed}. For the purposes of
later brevity, it is convenient to define a second, post-papal, cumulative
mortality,
\begin{eqnarray} \label{66} 
\delta^*(t, T) & = & \left\{ 
\begin{array}{l}
\delta_1 t \\ 
\delta_2 \left[ t - (\tau_1 - \tau_2) \right] + \delta_1 \left( \tau_1 - \tau_2 \right) 
\end{array} \hspace{5mm} \mbox{for} \hspace{5mm} 
\begin{array}{rcl}
&t&< \ \tau_1 - \tau_2 \\ 
&t&\ge \ \tau_1 - \tau_2,
\end{array} 
\right. \nonumber
\end{eqnarray}
one which commences at eclosion. 

\begin{table}[H]
\begin{center}
\begin{tabular}{c|c c c c c c c c}  
&  &  &  &  &  &  &  &  \\
T & \ $16 \ ^\circ\mathrm{C}$ & $18 \ ^\circ\mathrm{C}$ & $20 \ ^\circ\mathrm{C}$ & $22 \ ^\circ\mathrm{C}$ & $24 \ ^\circ\mathrm{C}$ & $26 \ ^\circ\mathrm{C}$ & $28 \ ^\circ\mathrm{C}$ & $30 \ ^\circ\mathrm{C}$ \\ \\ \hline \\
%N \ & \ $8 \times 10^6$ & $8 \times 10^6$ & $8 \times 10^6$ & $8 \times 10^6$ & $8 \times 10^6$ & $8 \times 10^6$ & $8 \times 10^6$ & $8 \times 10^6$ \\ \\
$\beta$ / \Female $^{-1}$ $\mathrm{day}^{-1}$ & \ 0.0082 & 0.0126 & 0.0167 & 0.0202 & 0.0233 & 0.0260 & 0.0284 & 0.0307 \\ \\
$\delta_0$ / $\mathrm{day}^{-1}$ \ & \ 0.01 & 0.01 & 0.01 & 0.01 & 0.01 & 0.01 & 0.01 & 0.01 \\ \\
$\delta_1$ / $\mathrm{day}^{-1}$ \ & \ 0.0188 & 0.0274 & 0.0352 & 0.0420 & 0.0480 & 0.0534 & 0.0584 & 0.0630 \\ \\
$\delta_2$ / $\mathrm{day}^{-1}$ & \ 0.0094 & 0.0137 & 0.0176 & 0.0210 & 0.0240 & 0.0267 & 0.0292 & 0.0315 \\ \\
\end{tabular}
\caption{Model parameters at different temperatures. $\beta$ is a maximum
possible eclosion (`birth') rate per female, $\delta_0$ is the pupal mortality,
$\delta_1$ is the pre-ovulatory mortality and $\delta_2$ is the adult mortality.} \label{parametersUsed}
\end{center}
\end{table}

\subsection{Eclosion Rate}

The eclosion rates in Table \ref{parametersUsed} were calculated on the basis of
the interlarval periods and a conservative estimate of pupal mortality (a
constant 1 \% \ $\mathrm{day}^{-1}$ for the puparial duration). This approach
could be problematic in the sense that pupal mortality is known to be density
dependent (\mbox{Du Toit}\nocite{DuToit}, 1954, and Rogers and
Randolph\nocite{RogersAndRandolph1}, 1990) and the pupal population can be
expected to shrink rapidly once spraying has commenced. 

Can the Table \ref{parametersUsed} values for $\beta$ therefore be reconciled
with the maximum eclosion rates one might expect to prevail at very low pupal
densities? After all, it is at these levels that the value must not be
underestimated. At 22$^\circ$C, for example, each female can only produce four pupae based on the 49-day, average, adult life-span of Glasgow\nocite{Glasgow1}, 1963. The absolute maximum that the population could
grow by, in the absence of any early mortality, would therefore be \mbox{2.8 \%
\Female $^{-1}$ $\mathrm{day}^{-1}$}. Of course, in the real world, a certain
amount of natural mortality is inevitable, no matter how low the pupal density.
One might therefore conclude that, if the Table \ref{parametersUsed} value of
2.02 \% \Female $^{-1}$ $\mathrm{day}^{-1}$ is wrong at low population
densities, it is not wrong by much. Some comfort can also be taken from evidence
which suggests that the pupa is only vulnerable for a more limited period and
that most of this density-dependent mortality is suffered earlier, rather than
uniformly, throughout the pupal stage (Parker\nocite{Parker1}, 2008). There
might also be a time lag associated with some forms of density dependence. In
either case, it would mean that the entire, pre-spray-deposited, pupal mass
should suffer a very similar, in appearance density-independent, mortality,
hence a constant eclosion rate. 

\section{Strategy for the Aerial Application of Insecticide} \label{strategy}

The reproductive life cycle of the tsetse fly makes it especially well disposed
to control through the cyclical application of insecticide. This is particularly
so in instances in which the blanket-application of a cycle is close to
instantaneous i.e. accomplished in a single night. Deltamethrin and endosulfan
are the usual insecticides of choice. The aerial application of small amounts of
deltamethrin can be so effective as to produce a mortality well exceeding 90\%,
under favourable circumstances.

\subsection{The Cycle Length}

Pupae present in the ground are unaffected by insecticide. The idea is therefore
to schedule follow-up operations shortly before the first flies to eclode,
after spraying, themselves mature and become reproductive. For temperatures of
$22 \ ^\circ\mathrm{C}$, or lower, both Jackson's curve and the data reported in
Hargrove\nocite{Hargrove3} (2004) suggest that spraying two days before the
first interlarval period (the one predicted using the
Hargrove\nocite{Hargrove4}, 1994 and 1995\nocite{Hargrove5}, coefficients) is
sufficient to ensure that none of the recently ecloded female flies ever give
birth prior to being sprayed. This observation is supported by the success of
operations such as those of Kgori et al.\nocite{Torr1}. For the present, the
two-day safety margin will be assumed to be 100\% effective at all
temperatures. Subsequent sprays are consequently scheduled two days short of the
first interlarval period, in other words,
\begin{eqnarray} \label{1}
\sigma = \tau_1(T) - 2,  
\end{eqnarray} 
in which $\sigma$ is the length of the interval between spray cycles and
$\tau_1$ is the relevant first interlarval period at the temperatures prevailing
for that interval. Table \ref{sigmaUsed} lists the values of $\sigma$ at
different temperatures. 

\begin{table}[H]
\begin{center}\begin{tabular}{c|c c c c c c c c}  
&  &  &  &  &  &  &  &  \\
T & \ $16 \ ^\circ\mathrm{C}$ & $18 \ ^\circ\mathrm{C}$ & $20 \ ^\circ\mathrm{C}$ & $22 \ ^\circ\mathrm{C}$ & $24 \ ^\circ\mathrm{C}$ & $26 \ ^\circ\mathrm{C}$ & $28 \ ^\circ\mathrm{C}$ & $30 \ ^\circ\mathrm{C}$ \\ \\ \hline \\
$\tau_1$ / $\mathrm{days}$ & 22 & 20 & 19 & 18 & 16 & 15 & 14 & 14 \\ \\
$\sigma$ / $\mathrm{days}$ & 20 & 18 & 17 & 16 & 14 & 13 & 12 & 12 \\ \\
$s$ & 7 & 6 & 5 & 5 & 4 & 4 & 4 & 4 \\ \\
\end{tabular}
\caption{The length of a cycle, $\sigma$, and the number of cycles, $s$, at different temperatures.} \label{sigmaUsed}
\end{center}
\end{table}

\subsection{The Number of Cycles}

Strategy dictates that spray cycles should be repeated until after the last
pre-spray-deposited pupae eclode. It is safer to continue until at least two
sprays after the emergence of the last flies from pre-spray-deposited pupae.
Given this precaution, it is difficult to imagine a scenario above $18 \
^\circ\mathrm{C}$ in which pre-spray-deposited pupae continue to eclode, in
any significant numbers, for more than a full spray cycle longer than the mean.
Hargrove\nocite{Hargrove3} (2004) determined that even the temperatures in
ant-bear burrows were, on average, only $2.2 \ ^\circ\mathrm{C}$ lower than the
ambient temperature. The number of cycles may therefore be formulated as follows.
\begin{eqnarray} \label{25}
\sum_{ i = 1 }^{ s - 2 } \sigma_i > \tau_0(T), 
\end{eqnarray} 
where $\sigma_i$ denotes the length of the $i$th interval between the $s$ total
spray cycles and $\tau_0(T)$ is the relevant puparial duration at the
temperatures prevailing for that interval. The total duration of the entire
spraying operation is $s - 1$ cycles and the time to the second last spray is
therefore $s - 2$ cycles. Expanding $\sigma_i$,
\begin{eqnarray} \label{27}
\sum_{ i = 1 }^{ s - 2 } [\tau_1(T) - 2]_i > \tau_0(T), 
\end{eqnarray} 
which at constant temperature yields
\begin{eqnarray} \label{30}
s > \frac{\tau_0}{\tau_1 - 2} + 2 \hspace{10mm} s \in \mathbb{Z}. 
\end{eqnarray} 
Table \ref{sigmaUsed} lists the values of $s$ at different temperatures.

At this early juncture it is worth noting that, at lower temperatures, the
required number of spray cycles is high, whereas at higher temperatures, the
required number of spray cycles is low. Note that although the values of $\beta$ and $\tau_0$ give rise to a greater number of pre-spray-deposited pupae at low temperature, it is the relative values of $\tau_0$ and $\sigma$ which ultimately have the most profound consequences for spraying at low temperature.

Cooler weather is, nonetheless, preferred for aerial spraying from a point of
view of spray efficacy (Hargrove\nocite{Hargrove11}, 2009). Very high kill rates
usually (though not always) come about as a result of the sinking air associated
with cooler weather. It favours the settling of insecticidal droplets. (Although
\mbox{Du Toit}\nocite{DuToit} (1954) makes mention of the sustained down draught
from a slow-moving helicopter, there are obviously distinct disadvantages to
such a method of insecticide application.) The effects of temperature on spray
efficacy are not modelled. For that matter, neither are the effects of anabatic
winds, nor the protection afforded by the forest canopy and multifarious other
variables relevant to spray efficacy. Spray efficacy is usually measured in the
field, with hindsight, rather than predicted. Three levels of spray efficacy are
entertained in this work. The kill rates of 99\%, 99.9\% and 99.99\% should be
thought of as being broadly associated with the warmer, intermediate and cooler
parts of the low-temperature range respectively. 

\section{Algorithm} \label{algorithm}

\begin{figure}[H]
    \begin{center}
\includegraphics[height=14cm, angle=0, clip = true]{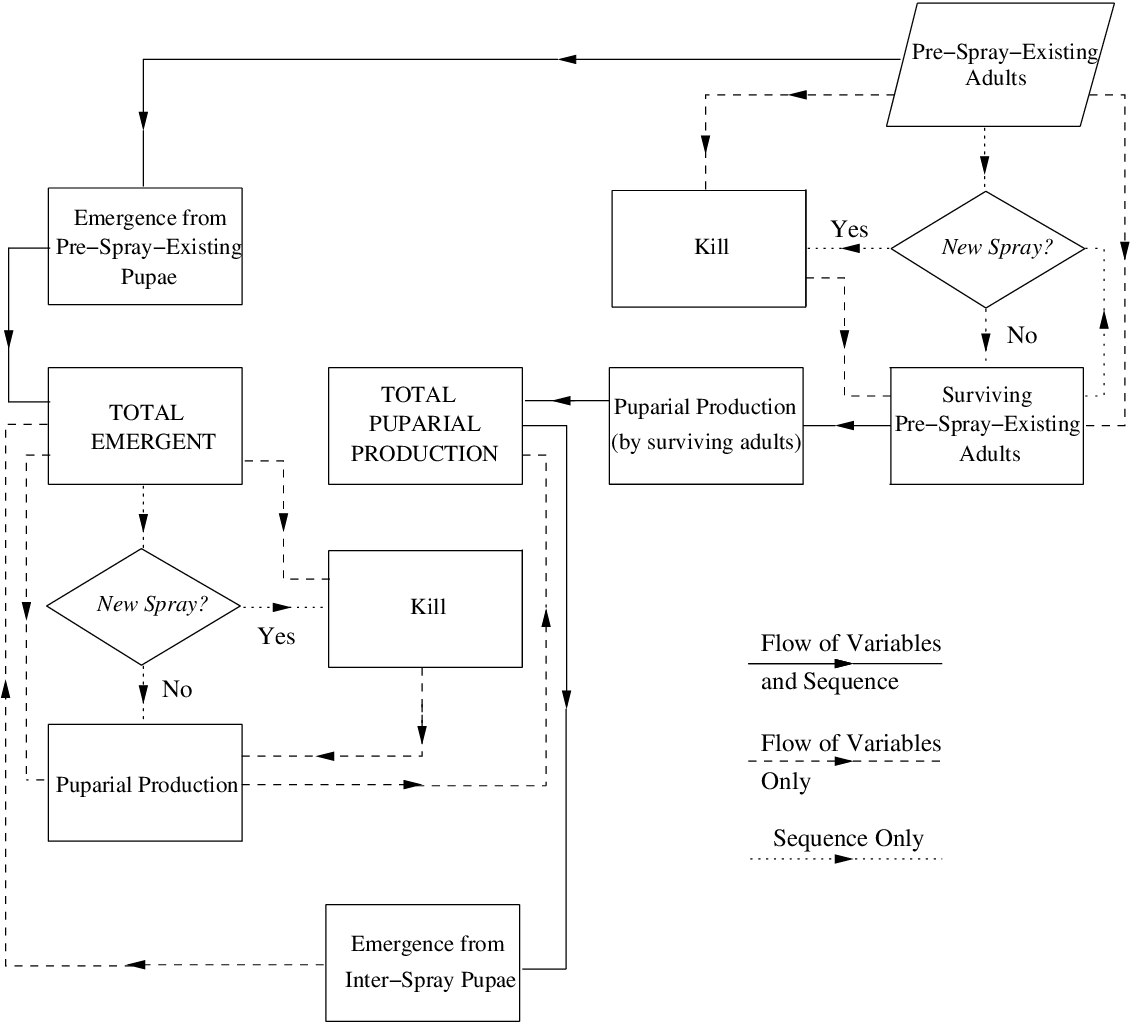}
\caption{Flow chart of the computation for a generalised day $i$.} \label{flowChart}
   \end{center}
\end{figure}

The data and formulae in Sections \ref{lifeCycle} and \ref{strategy} are sufficient information with which to model the performance of a worst-case,  tsetse population under conditions of aerial spraying. The formulae for the various durations are extended to accomodate variable temperature by averaging the values predicted for the daily temperature data. Figure \ref{flowChart} outlines the essentials of the algorithm.

\subsection{A Simple Test}

At $24 \ ^\circ\mathrm{C}$, four sprays (which define three spray cycles) are
required. The time to the second last spray is, for all practical purposes, one
puparial duration. The aerial spraying scenario at $24 \ ^\circ\mathrm{C}$ is
thus sufficiently simplified to lend itself to manual calculation. All the pupae
deposited during the first spray cycle eclode during the last spray cycle and
all the pupae still in the ground at the end of the operation were deposited
during the second and third spray cycles.  

\subsubsection*{Surviving Flies}

The pupae deposited during the first spray cycle are all descended from the
original, pre-spray-existing flies which survived the first spray. These pupae
eclode for the duration of the last spray cycle. The only other significant
contribution to the surviving fly population arises as a result of
pre-spray-deposited pupae, which eclode for the duration of the second cycle
and manage to survive the last two sprays. The outcome for surviving flies can
therefore be crudely formulated as
\begin{eqnarray*}
\sigma \frac{N}{\tau_2} e^{-\frac{\sigma}{2} \delta_2} \gamma e^{-\tau_0 \delta_0} e^{-\delta^*(\frac{\sigma}{2})} \phi^2
\ + \ \sigma \gamma N \beta e^{-\delta^*(\frac{\sigma}{2})} e^{-\sigma \delta_2} \phi^2 \ + \ O(\phi^3),&&
\end{eqnarray*}
in which mortalities for an average period of $\frac{\sigma}{2}$ have been substituted for those which really prevail for various portions of the spray cycle, $N$ is the original, steady-state, equilibrium number of females and $\phi$ is the probability of a fly surviving one spray.

\subsubsection*{Surviving Pupae}

There is only one significant contribution to the pupal population during the
second and third cycles. This is made by flies which eclode from
pre-spray-deposited pupae during the first and second cycles and which
subsequently survive a single spray to larviposit in the second and third
cycles. The pupal outcome can therefore be crudely formulated as
\begin{eqnarray*}
2 \sigma \gamma^2 N \beta e^{-\tau_0 \delta_0} e^{-\delta^*(\tau_1)} \phi \ + \ O(\phi^2).&&
\end{eqnarray*}
The results in Figures \ref{fliesPoint01}, \ref{fliesPoint001} and \ref{fliesPoint0001} compare very favourably with the values in Table \ref{handEstimatedResults} and this augurs well for the algorithm.

\begin{table}[H]
\begin{center}\begin{tabular}{c|c c c c}  
&  &  &  & \\
$\phi$ & \ flies & $\mathop{log}($ flies $)$ & pupae & $\mathop{log}($ pupae $)$ \\ \\ \hline \\
0.01 \ & \ 322 & 2.51 & 5670 & 3.75 \\ \\
0.001 \ & \ 3 & 0.51 & 567 & 2.75 \\ \\
0.0001 \ & \ 0 & - & 57 & 1.75 \\ \\
\end{tabular}
\caption{Estimated female survival for the simple case of $24 \ ^\circ\mathrm{C}$.} \label{handEstimatedResults}
\end{center}
\end{table}

\section{Results}

Results were generated for a range of temperatures and three different spray
efficacies using the algorithm briefly outlined in Section \ref{algorithm}. The
number of flies and pupae surviving an aerial spray operation involving an
initial, female, fly population of $8 \times 10^6$ are presented in Figures
\ref{fliesPoint01}, \ref{fliesPoint001} and \ref{fliesPoint0001}. For an area
measuring $8000 \mathrm{km}^2$, such a uniformly distributed population
corresponds to a population density of 1000\Female $\mathrm{km}^{-2}$. These are
more or less the typical characteristics of the kind of problem one expects in the vicinity of the Hluhluwe-iMfolozi reserve, although exactly how this value is arrived at requires some explanation. 

A recent study by Motloang et al.\nocite{MotloangMasumuVanDenBosscheMajiwaLatif}
(2009) brings the vector competence of {\em G. brevipalpis} into doubt. It
strongly suggests that {\em G. austeni} alone is the vector of trypanosomiasis
and that it is a highly competent one, at that. The Hell's Gate tsetse
population density was estimated to be at around 5000\Female $\mathrm{km}^{-2}$,
at the time when the mark-release-recapture experiments of Kappmeier
Green\nocite{KappmeierGreen} (2002) were carried out, and this value is typical of what one expects in good tsetse habitat (Hargrove\nocite{Hargrove11},
2009). If one then uses it to calibrate the {\em G. austeni} distribution of
Hendrickx\nocite{Hendrickx} (2007), one arrives at an estimate of around
1000\Female $\mathrm{km}^{-2}$ for large parts of the southern habitat. Further
north, in KwaZulu-Natal, the population density is, of course, much higher. The
population density in the Kgori et al.\nocite{Torr1}, 2006, operation was also
much higher. At this stage, however, only absolute population size, not
population density, is relevant. The only loss of generality in the results
comes from rounding off numbers. The 1000\Female $\mathrm{km}^{-2}$ results are,
therefore, readily adapted to the more usual occurrence of a 5000\Female
$\mathrm{km}^{-2}$ population, once density becomes relevant (in the analysis of
the results, at the end). The result is simply assumed to apply to an area
measuring $1600 \mathrm{km}^2$, instead of $8000 \mathrm{km}^2$.

Constant temperature was the context in which the results to follow were
generated. The variation in daily temperature presented in
Hargrove\nocite{Hargrove1} (1990) suggests real-life scenarios in which mean
temperature varies little from day to day. Note that surviving flies themselves
are of no real consequence to the outcome of a spraying operation. In fact no
female tsetse flies are expected to survive beyond the last cycle at all, given
a kill rate of 99.99\%.

\subsection{The Dependence on Temperature}

\subsubsection*{A 99\% Kill Rate}

\begin{figure}[H]
    \begin{center}
\includegraphics[height=11.9cm, angle=-90, clip = true]{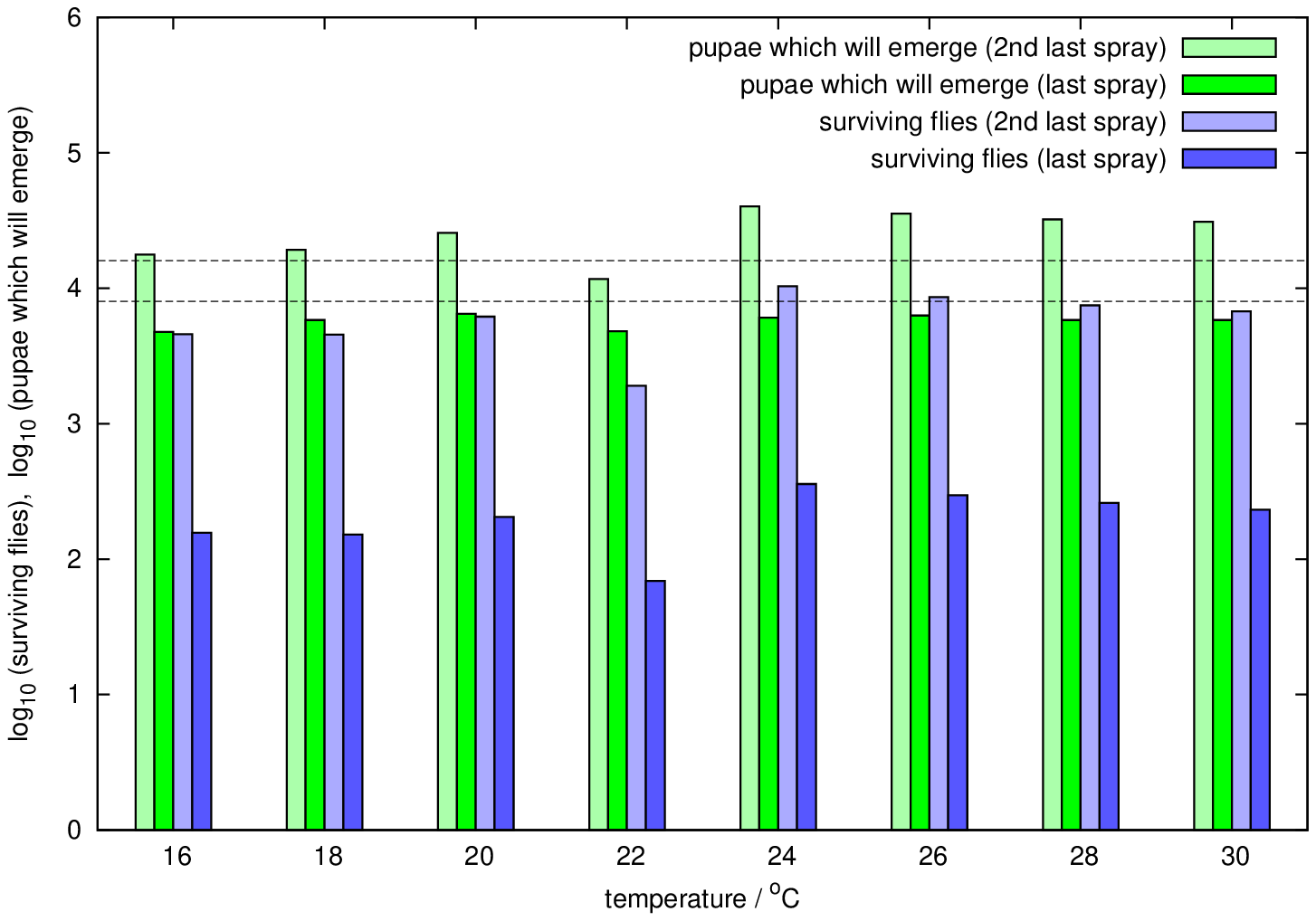}
\caption{Number of female tsetse flies and female pupae (which will survive to emerge) at the end of spraying, given a kill rate of 99\% and a starting population of $8 \times 10^6$ females. The dashed lines correspond to the 1\Female $\mathrm{km}^{-2}$ and 2\Female $\mathrm{km}^{-2}$ levels arising from a starting population density of 1000\Female $\mathrm{km}^{-2}$.} \label{fliesPoint01}
   \end{center}
\end{figure}

The results for a 99\% kill rate are presented in Figure \ref{fliesPoint01}. The results were generated for a range of constant temperatures.

\subsubsection*{A 99.9\% Kill Rate}

\begin{figure}[H]
    \begin{center}
\includegraphics[height=11.9cm, angle=-90, clip = true]{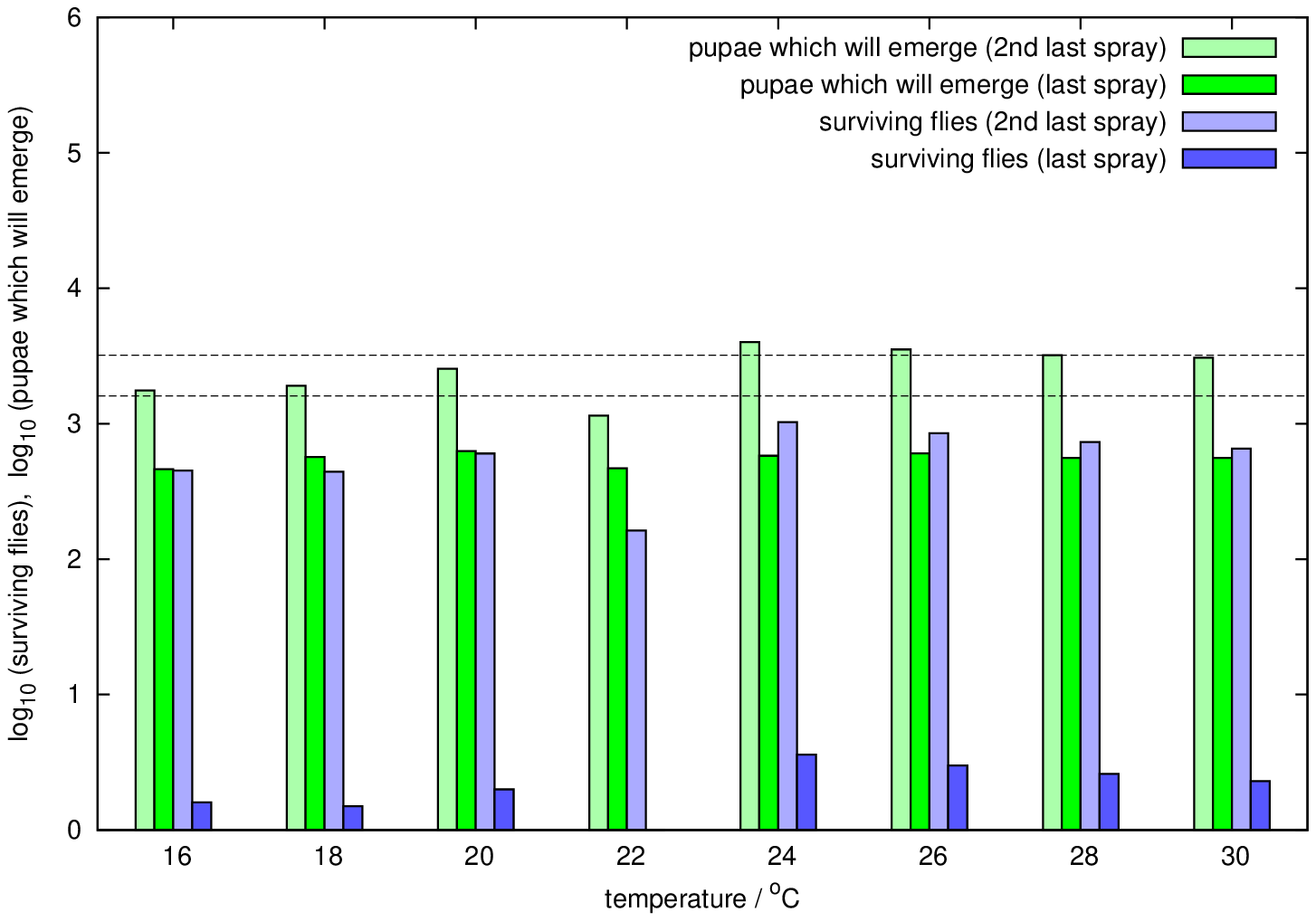}
\caption{Number of female tsetse flies and female pupae (which will survive to emerge) at the end of spraying, given a kill rate of 99.9\% and a starting population of $8 \times 10^6$ females. The dashed lines correspond to the 1\Female $\mathrm{km}^{-2}$ and 2\Female $\mathrm{km}^{-2}$ levels arising from a starting population of 5000\Female $\mathrm{km}^{-2}$.} \label{fliesPoint001}
   \end{center}
\end{figure}

The results for a 99.9\% kill rate are presented in Figure \ref{fliesPoint001}. The results were generated for a range of constant temperatures.

\subsubsection*{A 99.99\% Kill Rate}

\begin{figure}[H]
    \begin{center}
\includegraphics[height=11.9cm, angle=-90, clip = true]{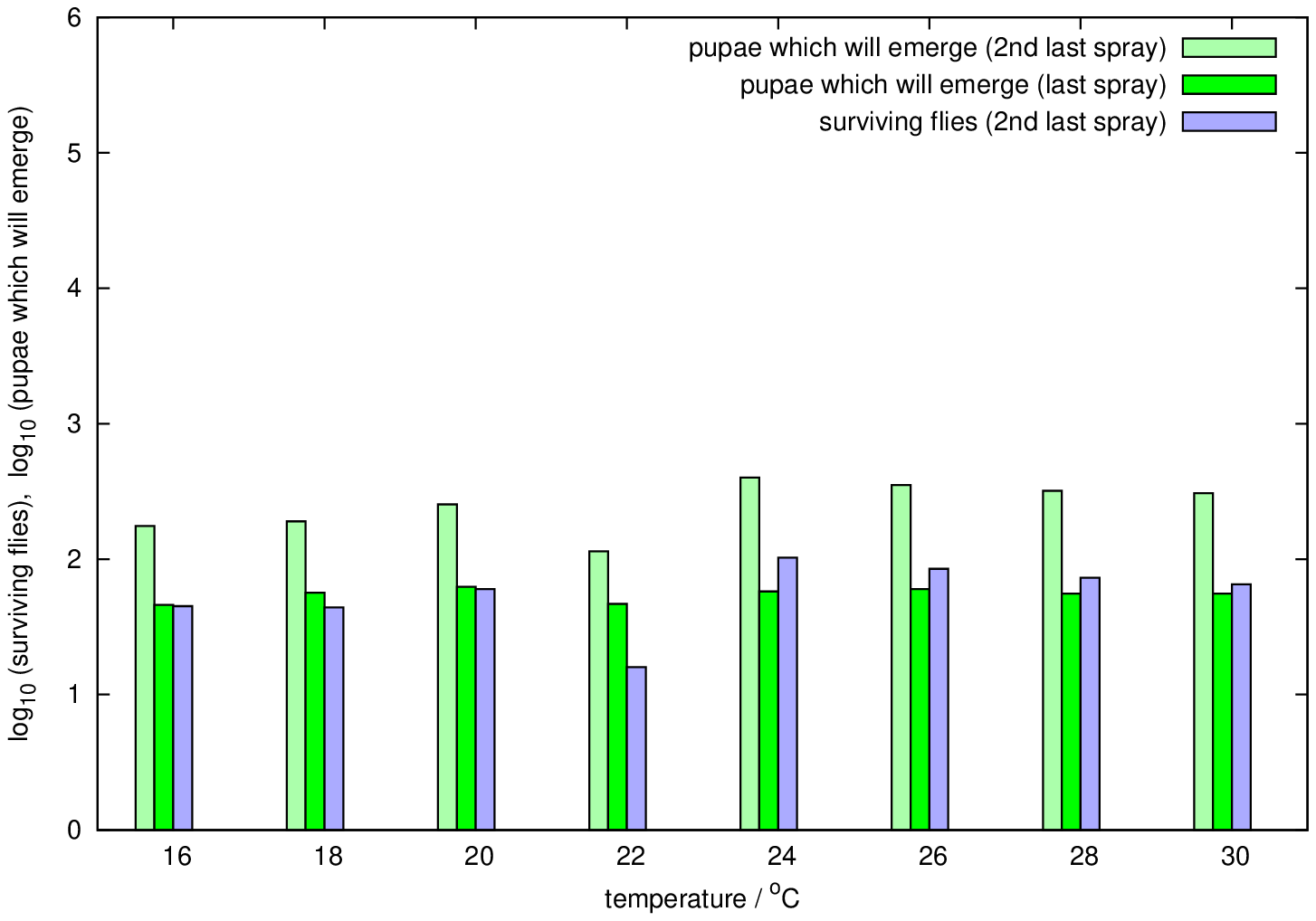}
\caption{Number of female pupae which will survive to emerge at the end of spraying, given a high kill rate of 99.99\% and a starting population of $8 \times 10^6$ females. No female flies survive.} \label{fliesPoint0001}
   \end{center}
\end{figure}

The results for a 99.99\% kill rate are presented in Figure \ref{fliesPoint0001}. The results were generated for a range of constant temperatures.

\subsection{The Origins of the Pupae Still to Emerge} \label{origins}

\begin{figure}[H]
    \begin{center}
\includegraphics[height=12.5cm, angle=-90, clip = true]{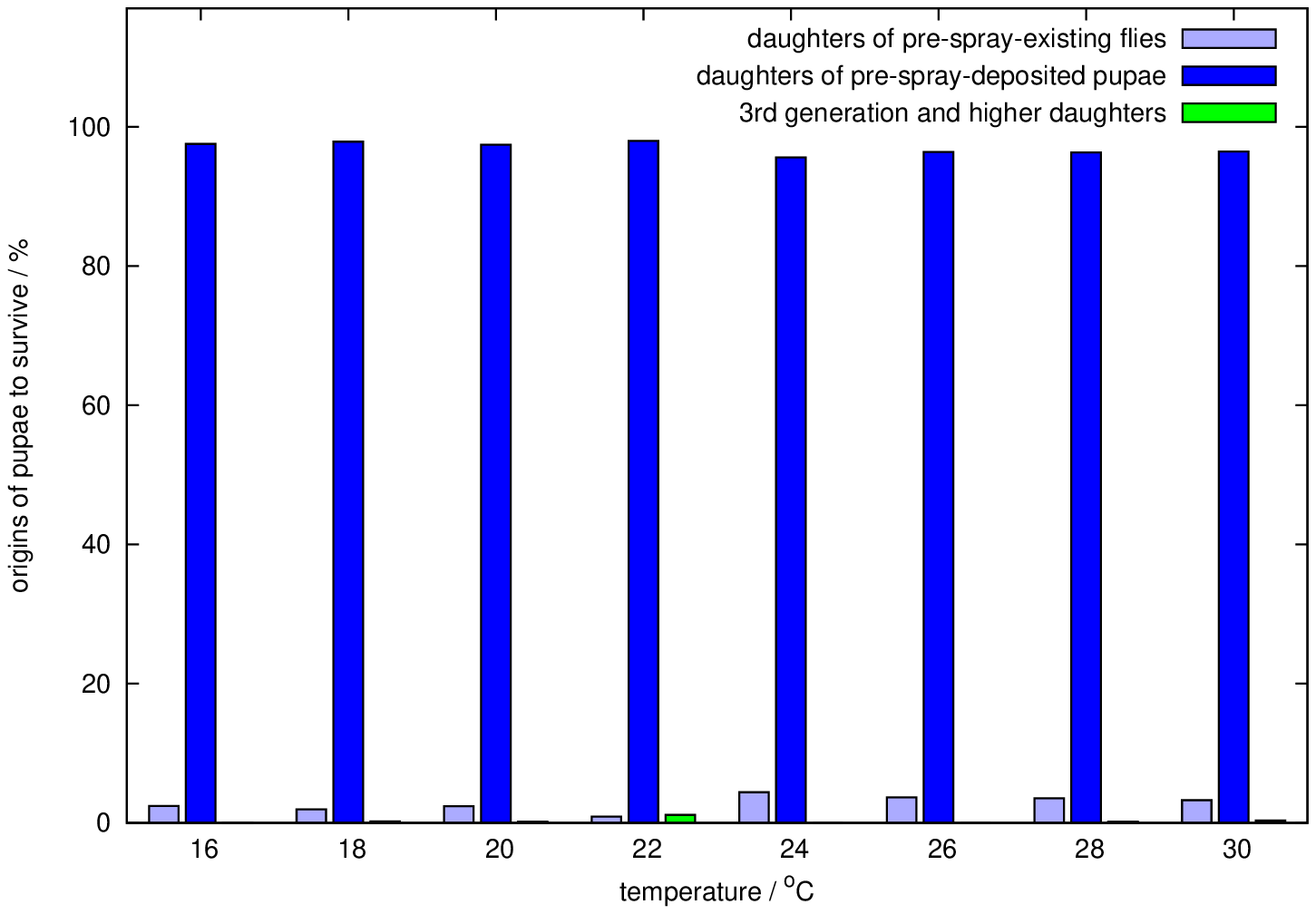}
\caption{A compositional analysis of the origins of the female pupae still in the ground at the end of spraying and which will survive to eclode, given a kill rate of 99\%.} \label{pupalCompositionPoint01}
   \end{center}
\end{figure}

A compositional analysis of the origins of female pupae still in the ground,
given a kill rate of 99\%, is presented in Figure \ref{pupalCompositionPoint01}.
Not only does dominance by the daughters-of-pre-spray-pupae category become
absolute at higher kill rates, the population of actual flies themselves becomes
vanishingly small. If, however, operations are halted one spray short, such a
predominance does not exist. Not only is the fly population still significant,
there is also a fairly large pupal population which is descended from the
original adults. 

\section{Interpretation of the Results}

The exact origins of the pupae in the ground at the end of spraying is a
question of considerable interest in its own right. The answer, nevertheless,
also provides an important clue for the elucidation of the overall results and
the discovery of certain key temperatures, just below which the spraying
operation achieves its maximum effect.

\subsection{Elucidating the Results}

Given that the daughters of pre-spray-deposited pupae are far and away the
greatest threat to success, then it stands to reason that the survival of their
immediate ancestors is also of key interest. A simple explanation of the results
becomes immediately apparent on considering the fate of pre-spray-deposited
pupae, during spraying. The explanation lies at the very heart of aerial
spraying strategy. From a metabolic point of view, spraying continues for longer
under certain conditions and the effects are twofold. 

\begin{figure}[H]
    \begin{center}
\includegraphics[height=12.5cm, angle=-90, clip = true]{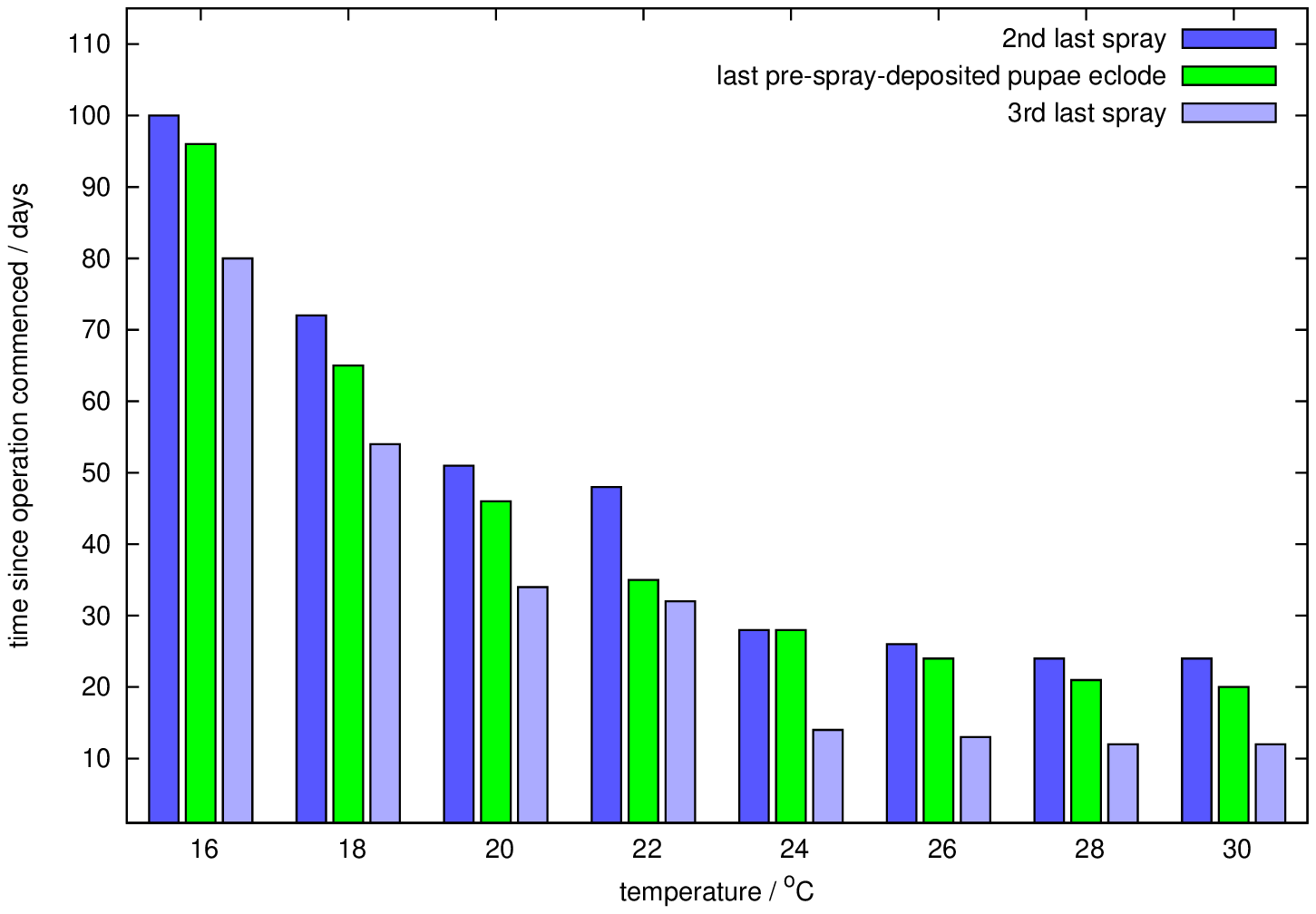}
\caption{The underlying explanation for the results.} \label{duration3rdLastAnd2ndLast}
   \end{center}
\end{figure} 

At such high kill rates it is not surprising that the primary influence on the
outcome of the spraying operation is, for flies, the proportion which were
subjected to only two sprays, instead of three. In other words, the outcome is
directly influenced by the fraction of the second last spray cycle for which
pre-spray-deposited pupae still eclode. At $22 \ ^\circ\mathrm{C}$, for
example, a very small fraction of the flies, which ecloded from
pre-spray-deposited pupae, are subjected to only the last two sprays (Figure
\ref{duration3rdLastAnd2ndLast}). At $24 \ ^\circ\mathrm{C}$ a very large
fraction of the flies, which ecloded from pre-spray-deposited pupae, are
subjected to only two sprays (Figure \ref{duration3rdLastAnd2ndLast}), hence the
jump in fly survival between $22 \ ^\circ\mathrm{C}$ and $24 \ ^\circ\mathrm{C}$
(Figure \ref{fliesPoint01}). The implications of these self-same circumstances
are just as important for pupae. Almost all the pupae deposited during the first
two spray cycles eclode during the operation, in time to be sprayed, instead of
only a portion of them. From a metabolic point of view, spraying continues for
longer. The first two cycles then also constitute a far greater portion of the
period during which most of the inter-spray larvae were deposited. Note that
both the aforementioned effects are obviously more pronounced when there are
fewer spray cycles i.e. at high temperature.

The above phenomena are believed to have been operative in the 2001 operation of
Kgori et al.\nocite{Torr1} (2006), which was terminated early (after what would
otherwise have been the second last spray). Almost all the pupae deposited
during their first cycle would have ecloded and been sprayed by the end of the
operation, furthermore, the duration of that first cycle constituted a
substantial portion of the duration of their operation (almost a quarter).
Secondly, only a very small fraction of the flies, which ecloded from
pre-spray-deposited pupae, were subjected to their last spray alone (what
would ordinarily have been the last two sprays). It is due to these facts that
the success of that 2001 operation can be attributed, in spite of them having
halted their operation one spray short.

\subsection{The Discovery of Key Temperatures}

\begin{figure}[H]
    \begin{center}
\includegraphics[height=12.5cm, angle=-90, clip = true]{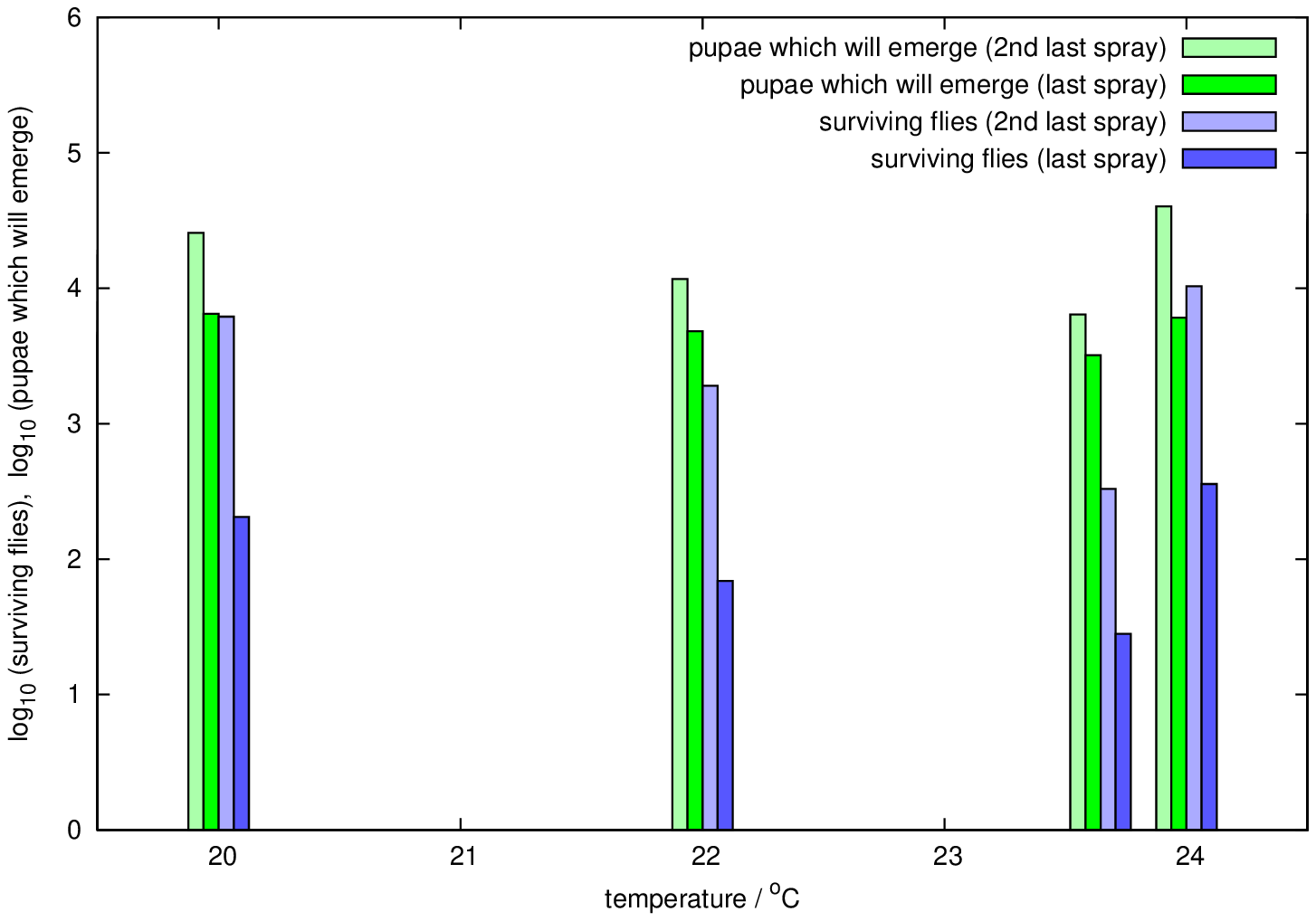}
\caption{The number of female tsetse flies and female pupae (which will survive
to emerge), given a kill rate of 99\%. As the mean temperature rises, so the
puparial duration shortens, rendering the last spray progressively less
relevant. By $23.645 \ ^\circ\mathrm{C}$, the fraction of flies which ecloded
from pre-spray-deposited pupae and which are subjected to only the last two
sprays, is minute. In addition, almost all the pupae deposited during the first
two spray cycles eclode during the operation, in time to be sprayed.} \label{log10MagicTemperaturesPoint01}
   \end{center}
\end{figure}

This leads to the discovery of what, in theory, are certain key temperatures. As
the mean temperature rises, so the puparial duration shortens, rendering the
last spray progressively less relevant. These are the temperatures at which the
length of the puparial duration approaches the time to the third last spray. For
such temperatures one obtains what amounts, effectively, to an extra spray. The
mean temperature eventually reaches a level where what was the last spray cycle
can be dispensed with. One obtains a favourable result just before strategy
(dictated by temperature) prescribes a reduction in the number of spray cycles
(Figure \ref{log10MagicTemperaturesPoint01}). There exist three, relevant
temperatures just below which very few of the pupae, deposited during the second
spray cycle, do not eclode in time for the last spray. At these same
temperatures, very few of the flies which eclode from pre-spray-deposited
pupae are subjected to only two sprays. Almost all are subjected to a minimum of
three sprays. Under these favourable circumstances the last spray is sometimes
omitted (as Kgori et al.\nocite{Torr1}, 2006, successfully did in their 2001
operation). Just above these temperatures the opposite is true. The last spray
cannot be abandoned. It is both relevant and necessary and an extra spray might
even be entertained. 

The temperatures at which strategy prescribes a reduction in the number of spray
cycles are found by solving the following equation:  
\begin{eqnarray} \label{29}
\sum_{i = 1 }^{ s - 3 } \left[\tau_1(T) - 2 \right]_i = \tau_0(T). 
\end{eqnarray}  
In this case, Newton's method was used to produce the values in
Table \ref{magicTemperatures}. The temperatures in Table \ref{magicTemperatures}
are limiting temperatures. Why the decimal places? Although the last spray
becomes progressively less relevant as the mean temperature increases, the exact
point at which it is finally deemed to be superfluous and abandoned is an
artefact of strategy; a man-made decision. The formula dictating the number of
spray cycles hinges on these decimal places. The mean temperature of the
environment is deemed to be either above or below one of the Table
\ref{magicTemperatures} temperatures and a decision is taken on that basis;
regardless of any natural variation or even error in measurement. (Obviously, if
one's prediction of the puparial duration is incorrect by a substantial fraction
of a cycle length, or one's mean temperature is incorrect to a magnitude of
degrees, this analysis will be rendered useless.) The existence of these
limiting temperatures has its origins in the fact that the formula for the
number of sprays is a discrete function of temperature, whilst that for the
puparial duration is a continuous one. Note that the limiting temperatures are
obviously irrelevant if one does not adhere to strategy, changing the number of
spray cycles when required to do so. One might, however, wish to modify that
strategy based on a better understanding of it.

\begin{table}[H]
\begin{center}\begin{tabular}{c|c c c}  
&  &  &  \\
$T + {\displaystyle \lim_{\epsilon \rightarrow 0^+}} \epsilon$ \ & \ $17.146 \ ^\circ\mathrm{C}$ & $19.278 \ ^\circ\mathrm{C}$ & $23.645 \ ^\circ\mathrm{C}$ \\ \\ \hline \\
$s$ & 7 & 6 & 5 \\ \\
$\sigma$ / $\mathrm{days}$ & 19.145 &  17.396 & 14.587 \\ \\
$\tau_0$ / $\mathrm{days}$ & 76.580 & 52.189 & 29.173 \\ \\
\end{tabular}
\caption{Aerial spraying strategies for temperatures at which the time to the third last spray cycle approaches the value of one puparial duration.} \label{magicTemperatures}
\end{center}
\end{table}

Notice, in Figure \ref{duration3rdLastAnd2ndLast}, that adhering to strategy
while spraying at any temperature above $23.645 \ ^\circ\mathrm{C}$ is not as
likely to be successful as below it and this fact is evident in the results. The
number of flies which eclode from pre-spray-deposited pupae to be subjected to
only two sprays will never be low and Figure \ref{pupalCompositionPoint01}
suggests a significant number of pupae from the second spray cycle have not yet
ecloded by the end of the operation. The number of sprays does not change again
above $23.645 \ ^\circ\mathrm{C}$. Above $23.645 \ ^\circ\mathrm{C}$ it might be
prudent to deviate from strategy by scheduling an extra spray.

\section{A constant-temperature Formula}

A moderately large set of formulae and recurrence relations, which predict the
entire outcome of an aerial spraying operation, can be derived for constant
temperature scenarios. As it transpires, only one, relatively simple formula
accounts for well over 90\% of the pupae still in the ground at the end of
spraying (given the findings of Section \ref{origins} and Figure
\ref{pupalCompositionPoint01}). This formula, furthermore, largely accounts for
the entire surviving tsetse population, given a kill rate of 99\%, or better.
A formula for the category `daughters of pre-spray-deposited pupae, which will
eclode after spraying' is presently derived.

\begin{figure}[H]
    \begin{center}
\includegraphics[height=4.5cm, angle=0, clip = true]{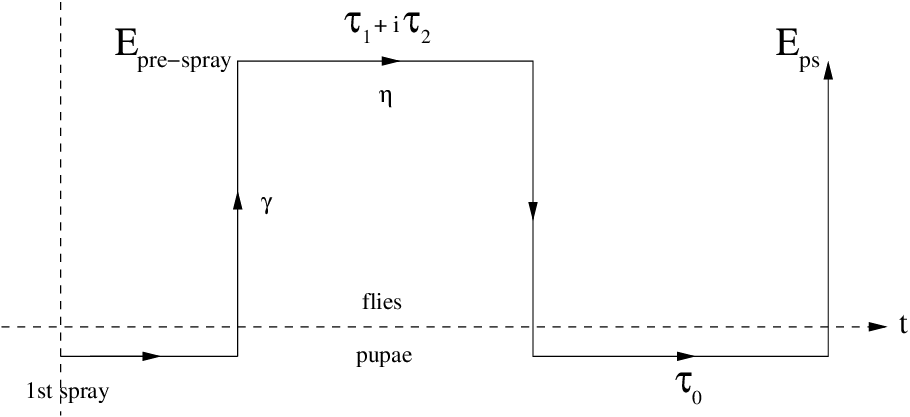}
\caption{Schematic diagram of second generation flies emerging from inter-spray pupae that are descended from pre-spray-deposited pupae.} \label{interSprayFromPreSpray}
   \end{center}
\end{figure}

The number of potential female parents, ecloding daily (for a limited period) from pre-spray-deposited pupae that will subsequently be inseminated, is
\begin{eqnarray} 
\gamma \eta E_{\mbox{\scriptsize pre-spray}} = \gamma \eta \beta N, \nonumber 
\end{eqnarray} 
in which $\beta$ is the steady-state, maximum possible, eclosion (`birth') rate
previously described, $N$ is the original, steady-state, equilibrium number of
females prior to spraying, $\gamma$ is the probability of being female and
$\eta$ is the probability of insemination. 

These pre-spray-deposited parents suffer a mortality of $\delta^*( \tau_1 + i
\tau_2, T )$ and are subjected to a total of  \begin{eqnarray*} \mbox{floor}
\left\{ \frac{ \breve{t} - \tau_0 }{\sigma} \right\} - \mbox{floor} \left\{
\frac{ \breve{t} - \tau_0 - \tau_1 - i \tau_2 }{\sigma} \right\}
\end{eqnarray*}   
spray cycles (by contemplating Figure \ref{interSprayFromPreSpray}) in which
$\breve t$ indicates the time-$\breve t$ cohort of their offspring. If a fly
survives one spraying cycle with probability $\phi$, then the probability that
it survives the above number of cycles is 
\begin{eqnarray*} 
&& \phi^{ \left( {\mbox{\scriptsize floor}} \left\{ \frac{ \breve{t} - \tau_0 }{\sigma} \right\} - {\mbox{\scriptsize floor}} \left\{ \frac{ \breve{t} - \tau_0 - \tau_1 - i \tau_2 }{\sigma} \right\} \right) }
\end{eqnarray*} 
(always assuming the probability of survival for each cycle is identical). Only an $e^{- \delta_0 \tau_0}$ fraction of their pupae survive to eclode.  

By contemplating Figure \ref{interSprayFromPreSpray}, the first and most obvious
requirement for second-generation descent from such parents, is a
restriction on the cohorts, $\breve{t}$. That is, \mbox{$\breve{t} > \tau_0 +
\tau_1$}. Secondly, only for a limited period of time (one puparial duration) do
parents originating from pre-spray-deposited pupae continue to emerge from the
ground. That is
\begin{eqnarray*} 
1 \ \le \ \breve{t} - \tau_0 - \tau_1 - i \tau_2 &\le& \tau_0  \hspace{10mm} i = 0, \ 1, \ \ldots, 
\end{eqnarray*} 
yielding a restriction on $i$,
\begin{eqnarray*}
i &\le& \mbox{floor} \left\{ \frac{1}{\tau_2} ( \breve{t} - \tau_0 - \tau_1 - 1 ) \right\}, \end{eqnarray*} 
and completing those on $\breve t$,
\begin{eqnarray*}
&& \tau_0 + \tau_1 \ < \ \breve{t} \ \le \ 2 \tau_0 + \tau_1 + i \tau_2. 
\end{eqnarray*} 
Collecting this information
\begin{eqnarray*} 
E_{ps}(\breve{t}) &=& \gamma \eta \beta N \sum_{i = 0}^{\mbox{\scriptsize floor}\left\{ \frac{1}{\tau_2} ( \breve{t} - \tau_0 - \tau_1 - 1 ) \right\} } \left[ e^{- \delta^*( \tau_1 + i \tau_2, T ) - \delta_0 \tau_0 } \frac{}{} \phi^{ \left( {\mbox{\scriptsize floor}} \left\{ \frac{ \breve{t} - \tau_0 }{\sigma} \right\} - {\mbox{\scriptsize floor}} \left\{ \frac{ \breve{t} - \tau_0 - \tau_1 - i \tau_2 }{\sigma} \right\} \right) } \right. \nonumber \\ 
&& \hspace{50mm} \left. \frac{}{} \left[ 1 - H( \breve{t} - 2 \tau_0 - \tau_1 - i \tau_2 ) \right] \ H( \breve{t} - \tau_0 - \tau_1 ) \right] \nonumber \\
\end{eqnarray*} 
in which $E_{ps}(\breve{t})$ is the time-$\breve{t}$ cohort (males and
females), immediately descended from pre-spray-deposited, female pupae, which
ecloded during the course of spraying, and $H$ is the version of the Heaviside
step function with $H(0) = 0$. 

The total number of such pupae which are both female and still in the ground at
the end of spraying, is the $\gamma$ fraction which will emerge between $\sigma
(s - 1) + 1$ and $\sigma (s - 1) + \tau_0$. That is,
\begin{eqnarray*} 
&& \hspace{-6mm} \eta \ \gamma^2 N \beta \sum_{{\breve t} = \sigma (s - 1) + 1}^{\sigma (s - 1) + \tau_0} \ \sum_{i = 0}^{ \mbox{\scriptsize floor}\left\{ \frac{1}{\tau_2} ( \breve{t} - \tau_0 - \tau_1 - 1 ) \right\} } \left[ e^{- \delta( \tau_0 + \tau_1 + i \tau_2, T ) } \phi^{ \left( {\mbox{\scriptsize floor}} \left\{ \frac{ \breve{t} - \tau_0 }{\sigma} \right\} - {\mbox{\scriptsize floor}} \left\{ \frac{ \breve{t} - \tau_0 - \tau_1 - i \tau_2 }{\sigma} \right\} \right) } \right. \hspace{10mm} \\
&& \hspace{65mm} \left. \frac{}{} \left[ 1 - H( \breve{t} - 2 \tau_0 - \tau_1 - i \tau_2 ) \right] \ H( \breve{t} - \tau_0 - \tau_1 ) \right]. \\
\end{eqnarray*} 
This formula is a good indicator of the entire outcome of an aerial spraying operation, given constant temperature and a kill rate of 99.9\%, or better. It accounts for well in excess of 90\% of the pupal population at a kill rate of 99\%.

\section{Creating the Closed Environment}

A cursory inspection of Hendrickx\nocite{Hendrickx} (2007) suggests that the
extant, forest-dwelling, tsetse populations of South Africa cannot be
considered closed and extend beyond its borders. The total extent of habitat is a further cause for concern. forest-dwelling species are notorious for their impartiality to odour-baited targets and some doubt has been expressed as to whether a barrier of the type used by Kgori et al.\nocite{Torr1} (2006) will be effective.

\begin{figure}[H]
    \begin{center}
\includegraphics[width=7.7cm, angle=0, clip = true]{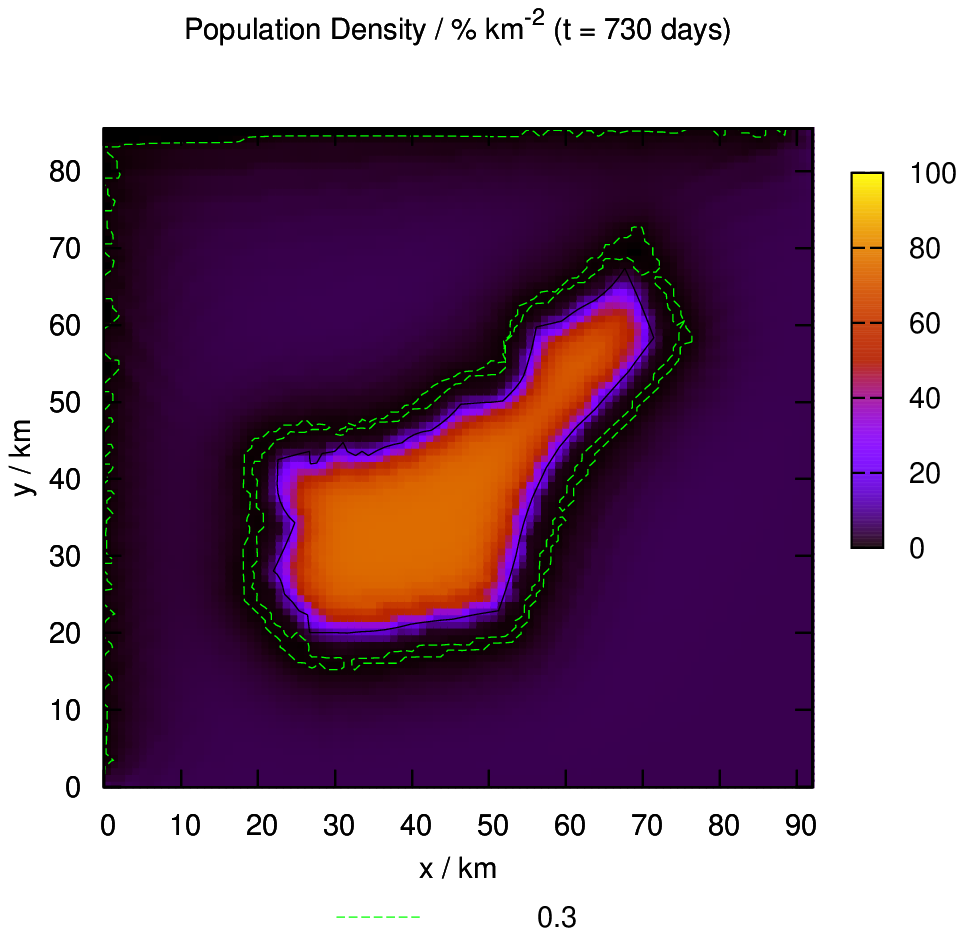}
\includegraphics[width=7.7cm, angle=0, clip = true]{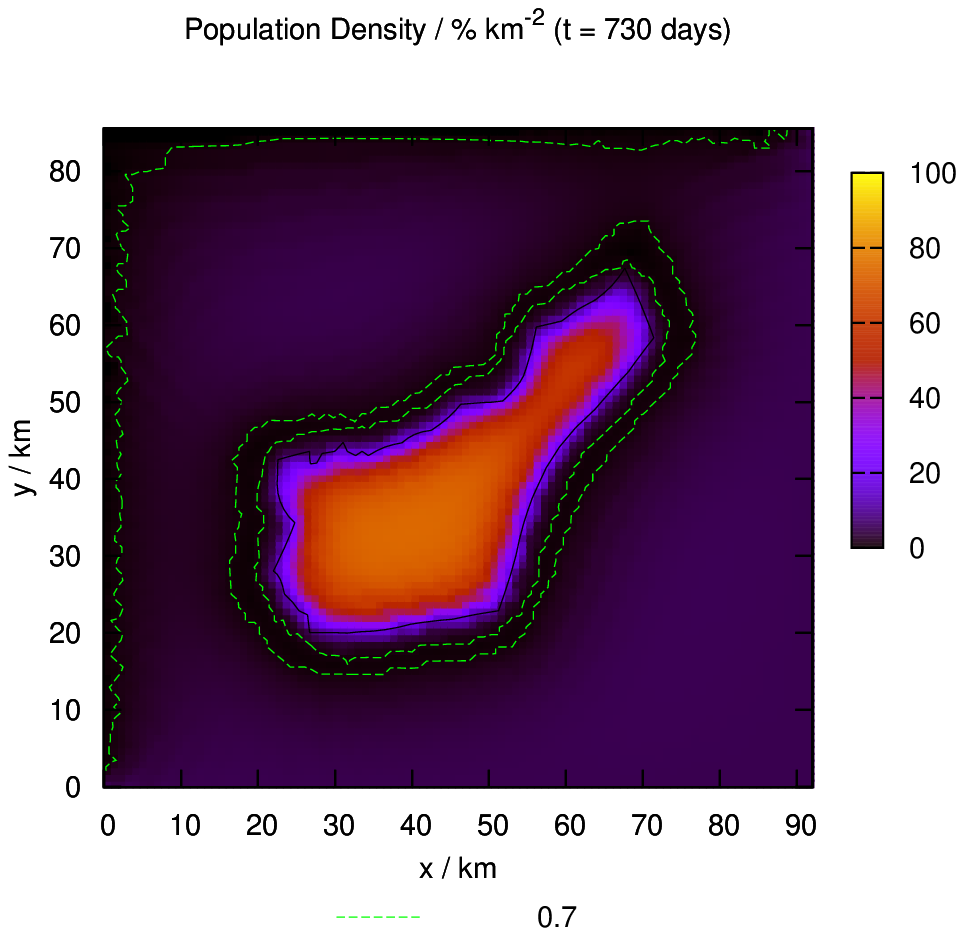}
\caption{The predicted effect of surrounding the Hluhluwe-iMfolozi Game Reserve
by an approximately 5 $\mathrm{km}$-wide barrier, with a \mbox{10 \%
$\mathrm{day}^{-1}$} mortality throughout. At left, a {\em G. austeni} diffusion
coefficient of \mbox{$0.04 \ \mathrm{km}^2 \ \mathrm{day}^{-1}$}. At right, a
worst-case, {\em G. austeni} diffusion coefficient of \mbox{$0.08 \
\mathrm{km}^2 \ \mathrm{day}^{-1}$}.} \label{isolatoryBarrier}
   \end{center}
\end{figure}

Childs\nocite{Childs3} (2010) and Esterhuizen et
al.\nocite{EsterhuizenKappmeierGreenNevillVanDenBossche} (2006) comprehensively
researched the design of such odour-baited, target barriers for {\em G.
austeni} and {\em G. brevipalpis}; albeit mostly from a point of view of a
control in its own right. The temporary barriers used for aerial spraying can be
made much wider than those used for containment, as there is no consequent waste
of the land they are intended to protect. Figure \ref{isolatoryBarrier} is the
predicted result of a more stringent, {\em G. austeni} isolation standard than
that used for control in Childs\nocite{Childs3}, 2010. It depicts the simulated
effect of surrounding the Hluhluwe-iMfolozi reserve by an approximately \mbox{5
$\mathrm{km}$-wide} barrier, throughout which there is a \mbox{10 \%
$\mathrm{day}^{-1}$} mortality. The simulation models migration as diffusion,
assumes growth is logistic and any artificially imposed mortality is modelled at
a constant rate (based on the posit of dispersal by random motion, attributed to
\mbox{Du Toit}\nocite{DuToit}, 1954, and others). It entertains two, different,
{\em G. austeni} diffusion coefficients, the higher one being a worst-case
coefficient. The \mbox{10 \% $\mathrm{day}^{-1}$} mortality, throughout that
barrier, corresponds to a target density of $33 \ \mathrm{km}^{-2}$ (this value
is not unreasonable when one considers that Kgori et al.\nocite{Torr1}, 2006,
used a target density of $16 \ \mathrm{km}^{-2}$ in a barrier which was five
times the width, in places). The effect of the Figure \ref{isolatoryBarrier}
barrier is to bring about a reduction in the {\em G. austeni} population, at the
interface of the reserve and the surrounding country side, by around two orders
of magnitude. A reduction by further orders of magnitude is best effected by
increasing the barrier's width, rather than the target density. 

The greater mobility of {\em G. brevipalpis} and its apparent impartiality to
odour-baited targets is cause for concern from the point of view of
containment, although mainly due to the fact that its susceptibility to
odour-baited targets was never really determined by Esterhuizen et
al.\nocite{EsterhuizenKappmeierGreenNevillVanDenBossche} (2006). Greater
mobility is a vulnerability, from a point of view of eradication by way of
odour-baited targets (Childs\nocite{Childs3}, 2010). Fortunately, Motloang et
al.\nocite{MotloangMasumuVanDenBosscheMajiwaLatif} (2009) have recently brought
{\em G. brevipalpis}' competence as a vector into question. 

\section{Conclusions}

Repeated spray cycles are scheduled at intervals two days short of the first
interlarval period and continue until two sprays subsequent to the eclosion of
the last, pre-spray-deposited, female pupae. Spray efficacy is found to come
at a price due to the greater number of cycles necessitated by cooler weather.
The greater number of cycles is a consequence of a larger ratio of puparial
duration to first interlarval period at lower temperatures. The prospect of a
more expensive spraying operation at low temperature, due to a greater,
requisite number of spray cycles is, however, one which is never confronted in
the real world. In reality, one has to strive towards the kill rates used in
this work and the only way such rates can be attained is by spraying at as low a
temperature as possible (Hargrove\nocite{Hargrove11}, 2009). 

Costs and the settling of insecticide droplets aside, this investigation
determines that there is little absolute difference between the outcomes of
aerial spraying at different temperatures. (Of course, it is the smallest of
differences which may ultimately determine the viability of any founding
population which survives spraying, as is evident from
Hargrove\nocite{Hargrove10}, 2005). The relative difference in the outcomes is,
however, significant and so any generalisable observations, or trends, which can
be made could therefore be profoundly relevant to the success of an aerial
spraying operation.  

The actual flies themselves (as distinct from pupae) which survive the last
spray are of no real consequence to the outcome of an aerial spraying operation.
Pupae, still in the ground at the end of spraying, are identified as the main
threat to successful control by aerial spraying. They are predominantly the
immediate descendants of pre-spray-deposited pupae, which ecloded during
spraying (and not third generation, or higher, pupae). The constant-temperature
formula for female flies ecloding from such pupae,
\begin{eqnarray*} 
&& \hspace{-6mm} \eta \ \gamma^2 N \beta \sum_{{\breve t} = \sigma (s - 1) + 1}^{\sigma (s - 1) + \tau_0} \ \sum_{i = 0}^{ \mbox{\scriptsize floor}\left\{ \frac{1}{\tau_2} ( \breve{t} - \tau_0 - \tau_1 - 1 ) \right\} } \left[ e^{- \delta( \tau_0 + \tau_1 + i \tau_2, T ) } \phi^{ \left( {\mbox{\scriptsize floor}} \left\{ \frac{ \breve{t} - \tau_0 }{\sigma} \right\} - {\mbox{\scriptsize floor}} \left\{ \frac{ \breve{t} - \tau_0 - \tau_1 - i \tau_2 }{\sigma} \right\} \right) } \right. \hspace{10mm} \\
&& \hspace{65mm} \left. \frac{}{} \left[ 1 - H( \breve{t} - 2 \tau_0 - \tau_1 - i \tau_2 ) \right] \ H( \breve{t} - \tau_0 - \tau_1 ) \right], \\
\end{eqnarray*} 
is therefore a good forecast of the outcome of a spraying operation, given a
kill rate of 99.9\% or better (other categories of pupae and flies themselves
can still constitute up to almost 10\% of the surviving population for a 99\%
kill rate). If, however, operations are halted one spray short, these
generalisations can not be made. Not only is the fly population still
significant, there is also a fairly large pupal population descended from the
original adults. 

Given the high kill rates attainable, it is not surprising that the outcome, for
flies (as distinct from pupae), is largely determined by the size of the
emergent population which was only subjected to the last two sprays (Figure
\ref{duration3rdLastAnd2ndLast}). At $22 \ ^\circ\mathrm{C}$, for example, a
very small fraction of the flies, which ecloded from pre-spray-deposited
pupae, are subjected to only the last two sprays. Most are subjected to at least
three sprays. At $24 \ ^\circ\mathrm{C}$, however, a very large fraction of the
flies, which ecloded from pre-spray-deposited pupae, are subjected to only two
sprays instead of three, hence the jump in fly survival between $22 \
^\circ\mathrm{C}$ and $24 \ ^\circ\mathrm{C}$ (Figure \ref{fliesPoint01}). The
implications of these self-same circumstances are just as important for
inter-spray pupae. Almost all the pupae deposited during the first two spray
cycles eclode during the operation, in time to be sprayed, instead of only a
smaller portion of them. Those first two cycles then also constitute a far
greater portion of the period during which most of the inter-spray larvae were
deposited. In summary, the additional effectiveness can largely be attributed to
the required number of sprays being close to borderline. From a metabolic point
of view, spraying continues for longer. This leads to the discovery of what, in
theory, are certain key temperatures; temperatures at which the time between the
first and third last sprays approaches one puparial duration. Aerial spraying
strategy is most effective against tsetse at temperatures just below either
$17.146 \ ^\circ\mathrm{C}$, $19.278 \ ^\circ\mathrm{C}$ or $23.645 \
^\circ\mathrm{C}$, in terms of the Hargrove\nocite{Hargrove3} (2004) formulae.
Conversely, the strategy is at its weakest if applied at temperatures only
fractionally above those three temperatures, $24 \ ^\circ\mathrm{C}$ in Figure
\ref{fliesPoint01} being a case in point. Spraying at any temperature above
$23.645 \ ^\circ\mathrm{C}$ is, in fact, a bad idea from the aforementioned
points of view. Just how well the prevailing temperature can be predicted for a
refinement in strategy is another question. Certainly, with hindsight, one can
base one's expectations and a decision to terminate an operation one spray short
on the disparity between the puparial duration and the time taken over the first
$s - 3$ spray cycles. A disparity of anything close to the length of a spray
cycle advocates caution (e.g. $24 \ ^\circ\mathrm{C}$ in Figures
\ref{duration3rdLastAnd2ndLast} and \ref{fliesPoint01}), whereas one which comes
close to vanishing should be interpreted as being auspicious (e.g. $22 \
^\circ\mathrm{C}$ in Figures \ref{duration3rdLastAnd2ndLast} and
\ref{fliesPoint01}). The effects of natural mortalities are all very small in comparison to those due to aerial spraying and they have little bearing on the overriding trends reported in this work

One would expect the pupal population to be at its most vulnerable at $16 \
^\circ\mathrm{C}$, given the cumulatively high, natural mortality which prevails
at this lower-temperature extreme of habitat (a consequence of a prolonged
puparial duration). For a given spray efficacy, the suggestion of this model is
that the surviving pupal population is only slightly diminished at $16 \
^\circ\mathrm{C}$. In reality, one might anticipate a more abrupt, as well as a
more profound, effect. To be fair, comparison is complicated very slightly by
the fact that the same equilibrium fly population at lower temperatures implies
a fractionally greater pupal mass in the ground than at higher temperatures. The
starting populations are not perfectly equivalent. One might therefore consider
including a factor of $10^{-0.044}$, when converting the $16 \ ^\circ\mathrm{C}$
results to population densities for comparison with the same for $30 \
^\circ\mathrm{C}$. On the other hand, pupal mortality at $30 \ ^\circ\mathrm{C}$
has been underestimated.   

A summary interpretation of Figures \ref{fliesPoint01} and \ref{fliesPoint001}
would be that a kill rate of 99\% will probably be adequate for the eradication
of a 1000\Female $\mathrm{km}^{-2}$ population, while a kill rate of 99.9\% is
likely to achieve the same end for one of 5000\Female $\mathrm{km}^{-2}$. One
can probably terminate the operation one spray short in the latter instance. If
one assumes an initial fly population density of 1000\Female $\mathrm{km}^{-2}$
and the subsequent survivors to be uniformly distributed throughout the area in
question, then the surviving population amounts to less than 1\Female
$\mathrm{km}^{-2}$ for all kill rates and temperatures considered (Figure
\ref{fliesPoint01}). Extinction is therefore very likely for any of the
temperatures and efficacies under consideration, given a closed population and a
likelihood of insemination $\le$ 0.1 (Hargrove\nocite{Hargrove10}, 2005). If,
however, the operation is terminated one spray short, the surviving population
density can be as high as around 6\Female $\mathrm{km}^{-2}$ for some of the
kill rates and temperatures considered (Figure \ref{fliesPoint01}). Extinction
is therefore very much less likely for an operation which is terminated one
spray short at $24 \ ^\circ\mathrm{C}$, for example. For a 99.9\% kill rate, the
population densities resulting from an operation which is terminated one spray
short are all less than 1\Female $\mathrm{km}^{-2}$. If, instead, one assumes an
initial fly population density of 5000\Female $\mathrm{km}^{-2}$, greater than
3\Female $\mathrm{km}^{-2}$ will subsequently eclode for a kill rate of 99\%,
unless at one of the key temperatures already mentioned. For a kill rate of
99.9\% (Figure \ref{fliesPoint001}), the surviving population amounts to less
than 1\Female $\mathrm{km}^{-2}$ for all temperatures under consideration.
Caution aside, the results suggest the last spray might even be omitted if the
operation is carried out below $23 \ ^\circ\mathrm{C}$. Lastly, given a kill
rate of 99.99\% (Figure \ref{fliesPoint0001}), the aerial spraying operation can
be terminated one spray short, regardless of the temperature and for either of
the population densities entertained above. The assumption of consistency in the
insecticide application to a uniformly distributed, starting population is
obviously sufficient to produce the uniformly distributed outcome desired. If,
however, the flies are not uniformly distributed, the probability of extinction
falls off rapidly with the size of any surviving, female, founding population as
well as with a higher probability of insemination. 

Great care therefore needs to be taken in the aerial application of the
insecticide. (In the pursuit of a cause for the failure of the Lambwe Valley
operation analysed by Turner and Brightwell\nocite{Turner1}, 1986, suspicion is
cast on the aerial application of the insecticide. Among other things, orogenic
lift is a well known cause of anabatic winds). Any oversights can ultimately
result in recolonization by the smallest of pioneer populations and G.P.S. puts
the modern operation at a distinct advantage in this regard. Neither the effects
of anabatic winds, nor the protection afforded by the forest canopy are known in
the context of the extant, forest-dwelling, tsetse populations of South Africa.
\mbox{Du Toit}\nocite{DuToit} (1954) mentions that at least some of this terrain
is inaccessible to fixed-wing aircraft and he observed that the quantity of an
atomized insecticide, destined not to penetrate the forest canopy, was simply
related to the density of the foliage. For these reasons, he favoured a
semi-gaseous, D.D.T. smoke, however, the terrain even presented a problem then,
in that the smoke would tend to collect in valleys, leaving the steep sides
uncovered. \mbox{Du Toit}\nocite{DuToit} (1954) also describes encountering what
one would imagine to be strong, anabatic winds e.g. in the northern sector of
the Hluhluwe-iMfolozi reserve. The fact that both forest-dwelling species
survived the twentieth century onslaught and endure to the present, speaks for
itself. Closed populations will need to be created by temporary barriers of
odour-baited targets (such as the one used successfully by Kgori et
al.\nocite{Torr1}, 2006) and not enough is known for the containment of {\em G.
brevipalpis}.

The determination of the longest puparial duration at breeding sites is of
crucial importance to the planning of any aerial spraying operation. So much so,
that suspicion is also cast on the estimated temperature of pupal sites, in the
pursuit of a cause for the failure of the Lambwe Valley operation (mountainous
terrain appears to lie to the south and breeding sites might therefore have
slipped into deep shadow, or have been cooled by drainage, evaporation or
catabatic winds). The risk of a late eclosion should only become a reality at
very low temperatures (below $18 \ ^\circ\mathrm{C}$), due to the exponential
lengthening of the puparial duration, and this risk becomes quantifiable if
knowledge of the environment is sufficient to allow a stochastic treatment. A
shorter than predicted first interlarval period is yet another, strong candidate
for a cause of failure. Failure would be the consequence of only a small
percentage of larviposition prior to the two-day safety margin on which the
spray cycle is based. Again, Hargrove\nocite{Hargrove10} (2005) quantifies the
dangers in allowing the smallest of founding populations to survive. The
puparial duration of {\em G. brevipalpis} is exceptionally long, to the extent
that the formula Equation \ref{31} may underestimate it by almost 20\%. Caution
should therefore be exercised in using Equation \ref{31} to determine the time
at which the last {\em G. brevipalpis}, pre-spray-deposited pupae eclode and
consequently, the required number of spray cycles dictated by Equation \ref{30}.
There is good reason to suspect that {\em G. austeni} could also be problematic
from the point of view of a shorter first interlarval period. A refinement of
the existing formulae for the puparial duration and the first interlarval period
might be prudent in the South African context of a sympatric {\em G.
brevipalpis}-{\em G. austeni}, tsetse population. This is especially given that
the two-day safety margin for repeat spraying could already be problematic at
high temperature. The resulting aerial spraying strategy would then be
formulated using a {\em G. brevipalpis} puparial duration and a {\em G. austeni}
first interlarval period in Equations \ref{27} or \ref{30} and Equation
\ref{29}. 
 
\begin{figure}[H]
    \begin{center}
\includegraphics[height=12.5cm, angle=-90, clip = true]{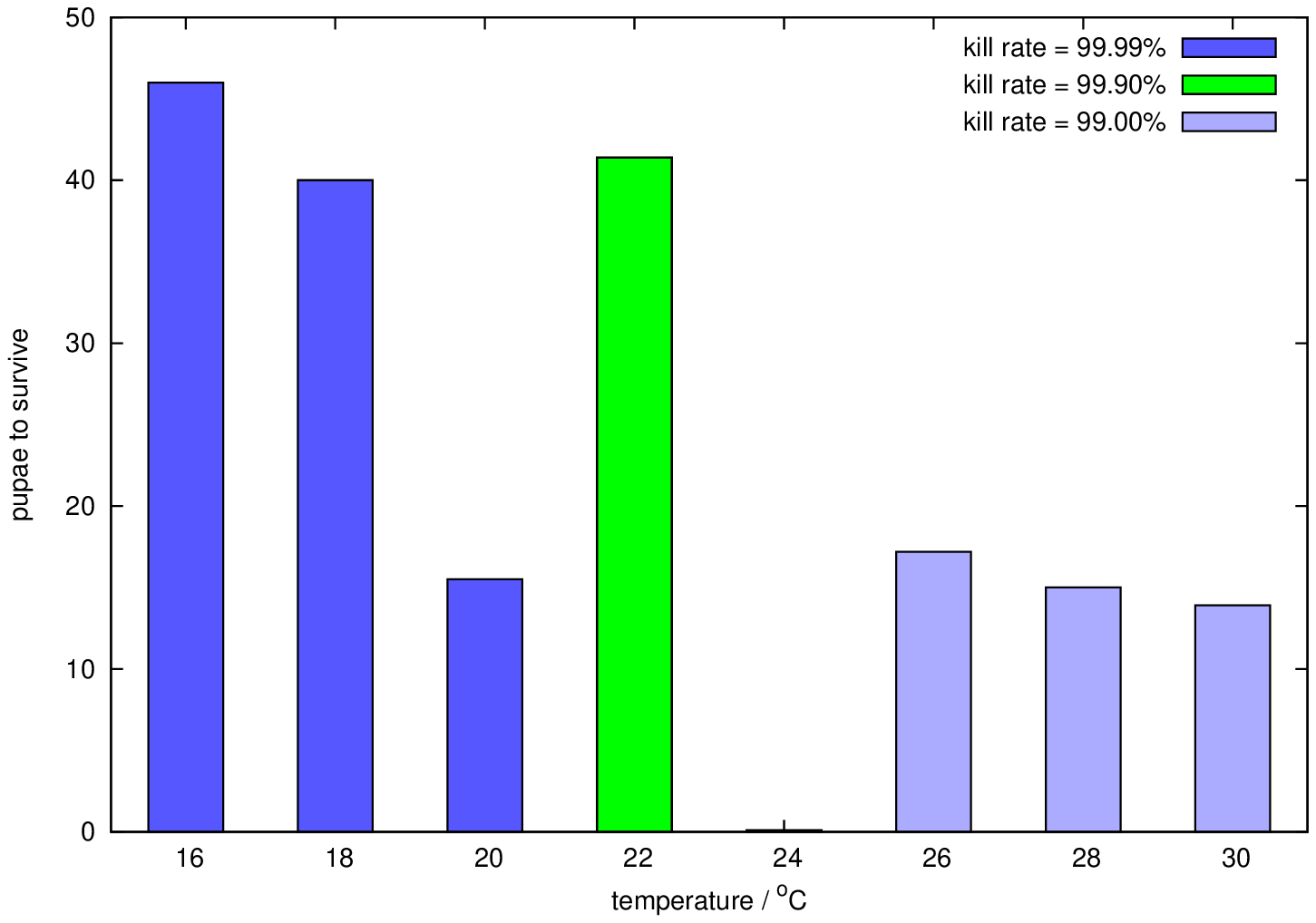}
\caption{A constant-cost (7 sprays) comparison of the numbers of pupae which will survive to emerge, assuming kill rates of 99.99\%, 99.9\% and 99\% can be associated with the low, intermediate and high temperature ranges respectively. (The implied variation of kill-rate with temperature is thought to be grossly understated.) The constant, 7-spray strategy obviously differs from that for which the results, in the other figures, were obtained.} \label{constantCostsPupae}
   \end{center}
\end{figure} 

The issue of kill rates and the associated cost is always going to be slightly
elusive due to circumstances unique to each and every aerial spraying operation.
It can, nevertheless, be resolved by argument to an extent. Consider comparing
the results for a constant number (and therefore constant cost) of spray cycles
for a range of temperatures. Suppose, for arguments sake, that the range in kill
rates considered in this work spans a $4 \ ^\circ\mathrm{C}$ range in
temperature. If the 99.99\% constant-cost results are attributed to the lowest
temperature, the 99.9\% constant-cost results to the intermediate temperature
and the 99\% constant-cost results to the highest temperature, the trend
favours spraying at low temperature (the data was considered too great a
digression for presentation). On this basis it could be argued that any extra
cost invested in spraying at low temperature is worthwhile. What if the range of
kill rates considered spans a bigger range in temperatures? Only when the kill
rates in question are assumed to span the entire range of temperatures
considered (Figure \ref{constantCostsPupae}) is there a slight reversal in trend
(certainly nothing of an order of magnitude). Only then can it be argued that
`you get what you pay for', regardless of the temperature. Those concerned with
the vagaries of spray efficacy are apt to point out that there is a very real
danger that you may not achieve anything near a 99\% kill rate at high
temperature. They would argue that a kill rate of 98\%, or 94\% (and, hence, a
failed operation) would be what one would expect. In reality the results become
progressively worse than in Figure \ref{constantCostsPupae} as the temperature
rises. At high temperature, there is also the possibility of some variance in
the first interlarval period to levels below the length of the spray cycle. The
range of kill rates used in this work only spans the lower end of the
temperature range investigated. In other words, taking efficacy, costs and the
reproductive life cycle of the tsetse fly into account, you get more than what
you pay for by conducting the aerial spraying operation at low temperature. A
simpler argument for spraying at low temperature is the prospect that one might
not otherwise succeed. Again, the question of kill rates and the associated cost
is, of course, really an academic one. In the real world kill rates are the
priority and one must therefore strive to spray at as low a temperature as
possible. 

Fortunately, there are already sufficient reasons to deter any real interest in
the high-temperature regime results. A cursory inspection of the data and
curves for the first interlarval period (in Hargrove\nocite{Hargrove3}, 2004)
suggests that, at around $26 \ ^\circ\mathrm{C}$ and above, there exists an ever
present danger implicit in a spray cycle based on a two-day safety margin: A
proportion of the flies, ecloding on the first day of a cycle, could just
succeed in depositing a pupa before being sprayed. Such pupae are, for all
cycles except the first (the last $s - 2$ cycles), destined to emerge only after
the completion of the operation. They will never be sprayed. Of these cycles, $s
- 3$ of them include eclosion from the very large, pre-spray-deposited pupal
mass. (No pre-spray-deposited pupae eclode on the first day of the last
cycle.) How does this translate into actual numbers? At $30 \ ^\circ\mathrm{C}$,
should a 10\% proportion have a first interlarval period less than two days
short of the mean, it would entail around $10^{3.8}$ additional, female pupae
in the ground; irrespective of any kill rate. At very high kill rates, the
implications are profound and the results, at $26 \ ^\circ\mathrm{C}$ and above,
should consequently be regarded with extreme caution. For such temperatures,
this work could well benefit from a stochastic treatment (although the outcome
is already foreseeable). Either that, or a change in strategy, such as a three-day safety margin.

The ethics and environmental impact of aerial spraying is really a topic in its
own right; one which many believe to have been sadly neglected for the most
part. Both deltamethrin and endosulfan are biodegradable. Maximum doses of 0.26
$\mathrm{g} \ \mathrm{ha}^{-1}$ to \mbox{0.3 $\mathrm{g} \ \mathrm{ha}^{-1}$}
deltamethrin were used in the landmark operation described by Kgori et
al.\nocite{Torr1} (2006). It is claimed that such doses have no, to little,
residual effect. Nagel\nocite{Nagel1} (1993), nonetheless, points out that shell
fisheries and other crustacean related enterprises probably need to be protected
from insecticide, as do honey bees. Some would argue that Nagel\nocite{Nagel1}
(1993) does not go far enough. They would argue that work, such as that by
Perkins and Ramberg\nocite{PerkinsAndRamberg} (2004), suggests that a full 10\%
of known species of shrimps and back-swimmers are permanently lost as a result
of aerial spraying. This is obviously a concern when it comes to habitat such as
the St. Lucia estuary, a world heritage site where one would anticipate a
certain amount of endemicity. Flight emergencies do happen and cattle died as a
result of one such incident where the insecticidal load had to be dumped, in
Zambia. Although Nagel\nocite{Nagel1} (1993) attributes these deaths to a
mismanaged response and non-compliance with subsequent advice, rather than the
insecticidal operation itself, it should be remembered that aerial spraying is
usually carried out in remote areas where there are often problems in
communication and levels of education are sometimes poor. A decisive
emergency-response strategy to implement precautions and clean-up operations
in the event of unforeseen accidents (e.g. spillages and flight emergencies)
therefore needs to be formulated. Many would argue in favour of more
discriminate means of control, such as pour-ons, dips and odour-baited
targets. They would also claim that if people are prepared to entertain concepts
such as S.I.T. (sterile insect technique), a bio-control programme utilizing
the tsetse fly's natural enemies, or other, competing, tsetse species which lack
the same vector competence, are all options which warrant investigation. A
bio-control method based, for example, on the ``spider islands'' of
Fiske\nocite{Fiske} (1920) has yet to be advanced, despite
Glasgow's\nocite{Glasgow1} (1963) view that {\em Nephila} had led to the
eradication of tsetse on Sumba Island circa 1947. One might also surmise that
such an intervention would have a greater environmental impact than the
deployment of odour-baited targets, although admittedly, one not as great as
that due to aerial spraying.   

\section{Acknowledgements} 

The original conception of this project was entirely John Hargrove's and it was under his close direction and generous guidance that the work was carried out at SACEMA, in Stellenbosch. SACEMA also funded the work in part. Abdalla Latif, Chantel De Beer and the Onderstepoort Veterinary Institute are thanked for their generosity in both facilitating and funding this work. The author is indebted to Andrew Parker for the information on puparial durations. Lastly, both anonymous reviewers and the editor are thanked for their suggestions and guidance. 

\nocite{Hargrove6}

\nocite{Hursey1}

\nocite{Hursey2}

\nocite{Randolph1}

\bibliography{aerialSprayingPaper}

\begin{thebibliography}{10}

\bibitem{Allsopp1}
R.~Allsopp.
\newblock Control of tsetse flies ({{\em {G}lossina}} spp) using insecticides:
  a review and future prospects.
\newblock {\em Bulletin of Entomological Research}, 75:1--23, 1984.

\bibitem{Anonymous}
Anonymous.
\newblock Notes for field studies of tsetse flies in {E}ast {A}frica.
\newblock Technical report, {E}ast {A}frica {H}igh {C}omission, {N}airobi,
  1955.

\bibitem{Childs2}
S.~J. Childs.
\newblock A model of pupal water loss in {{\em {G}lossina}}.
\newblock {\em Mathematical Biosciences}, 221:77--90, 2009.

\bibitem{Childs3}
S.~J. Childs.
\newblock The finite element implementation of a {K.P.P.} equation for the
  simulation of tsetse control measures in the vicinity of a game reserve.
\newblock {\em Mathematical Biosciences}, 227:29--43, 2010.

\bibitem{Chorley}
J.K. Chorley.
\newblock The bionomics of {{\em {G}}}{\em lossina morsitans} in the {U}mniati
  fly belt, {S}outhern {R}hodesia, 1922--1923.
\newblock {\em Bulletin of Entomological Research}, 20(3):279--301, 1929.

\bibitem{DuToit}
R.~Du~Toit.
\newblock Trypanosomiasis in {Z}ululand and the control of tsetse flies by
  chemical means.
\newblock {\em Onderstepoort Journal of Veterinary Research}, 26(3):317--387,
  1954.

\bibitem{EsterhuizenKappmeierGreenMarcottyVanDenBossche}
J.~Esterhuizen, K.~Kappmeier~Green, T.~Marcotty and P.~Van Den~Bossche.
\newblock Abundance and distribution of the tsetse flies, {{\em {G}}}{\em
  lossina austeni} and {{\em {G}}}{\em . brevipalpis} in {S}outh {A}frica.
\newblock {\em Medical and Veterinary Entomology}, 19(4):367--371, 2005.

\bibitem{EsterhuizenKappmeierGreenNevillVanDenBossche}
J.~Esterhuizen, K.~Kappmeier~Green, E.~M. Nevill and P.~Van Den~Bossche.
\newblock Selective use of odour--baited, insecticide--treated targets to
  control tsetse flies {{\em {G}}}{\em lossina austeni} and {{\em {G}}}{\em .
  brevipalpis} in {S}outh {A}frica.
\newblock {\em Medical and Veterinary Entomology}, 20:464--469, 2006.

\bibitem{EsterhuizenVanDenBossche}
J.~Esterhuizen and P.~Van Den~Bossche.
\newblock Protective netting, an additional method for the integrated control
  of livestock trypanosomiasis in {K}wa{Z}ulu-{N}atal {P}rovince, {S}outh
  {A}frica.
\newblock {\em Onderstepoort Journal of Veterinary Research}, 73(4):319--321,
  2006.

\bibitem{Fiske}
W.~F. Fiske.
\newblock Investigations into the bionomics of {{\em {G}}}{\em lossina
  palpalis}.
\newblock {\em Bulletin of Entomological Research}, 10:347--463, 1920.

\bibitem{Glasgow1}
J.~P. Glasgow.
\newblock {\em The Distribution and Abundance of Tsetse}.
\newblock International Series of Monographs on Pure and Applied Biology.
  Pergamon Press, 1963.

\bibitem{Hargrove1}
J.~W. Hargrove.
\newblock Age--dependent changes in the probabilities of survival and capture
  of the tsetse, glossina morsitans morsitans westwood.
\newblock {\em Insect Science and its Applications}, 11(3):323--330, 1990.

\bibitem{Hargrove2}
J.~W. Hargrove.
\newblock Age dependent sampling biases in tsetse flies (glossina). {P}roblems
  associated with estimating mortality from sample age distributions.
\newblock {\em Management of Insect Pests: Nuclear and Related Molecular and
  Genetic Techniques ***pp. International Atomic Energy Agency, Vienna.}, pages
  549--556, 1993.

\bibitem{Hargrove4}
J.~W. Hargrove.
\newblock Reproductive rates of tsetse flies in the field in {Z}imbabwe.
\newblock {\em Physiological Entomology}, 19:307--318, 1994.

\bibitem{Hargrove5}
J.~W. Hargrove.
\newblock Towards a general rule for estimating the day of pregnancy of field
  caught tsetse flies.
\newblock {\em Physiological Entomology}, 20:213--223, 1995.

\bibitem{Hargrove6}
J.~W. Hargrove.
\newblock Tsetse eradication; sufficiency, necessity and desirability.
\newblock Technical report, DFID Animal Health Programme, Centre for Tropical
  Veterinary Medicine, University of Edinburgh, 2003.

\bibitem{Hargrove3}
J.~W. Hargrove.
\newblock {\em The Trypanosomiases}.
\newblock Editors: I. Maudlin, P. H. Holmes and P. H. Miles. {CABI} publishing,
  {O}xford, U.K., 2004.

\bibitem{Hargrove10}
J.~W. Hargrove.
\newblock Extinction probabilities and times to extinction for populations of
  tsetse flies {{\em {G}}}{\em lossina} spp. ({D}iptera: Glossinidae) subjected
  to various control measures.
\newblock {\em Bulletin of Entomological Research}, 95:1--9, 2005.

\bibitem{Hargrove11}
J.~W. Hargrove.
\newblock {\em By communication}.
\newblock 2009.

\bibitem{HargroveAndWilliams}
J.~W. Hargrove and B.~G. Williams.
\newblock Optimized simulation as an aid to modelling, with an application to
  the study of tsetse flies, {{\em {G}lossina morsitans morsitans}} ({{\em
  {D}iptera: Glossinidae}}).
\newblock {\em Bulletin of Entomological Research}, 88:425--435, 1998.

\bibitem{Hendrickx}
Guy Hendrickx.
\newblock Tsetse in {K}wazulu {N}atal -- an update --.
\newblock Technical report, Agricultural and Veterinary Intelligence Analysis,
  2007.

\bibitem{Hursey1}
B.~S. Hursey and R.~Allsopp.
\newblock Sequential applications of low dosage aerosols from fixed wing
  aircraft as a means of eradicatiing tsetse flies ({{\em {G}lossina}} spp)
  from rugged terrain in zimbabwe.
\newblock Technical report, Harare, Tsetse and Trypanosomiasis Control Branch,
  1983.

\bibitem{Hursey2}
B.~S. Hursey and R.~Allsopp.
\newblock The eradication of tsetse flies ({{\em {G}lossina}} spp) from from
  western {Z}imbabwe by integrated aerial and ground spraying.
\newblock Technical report, Harare, Tsetse and Trypanosomiasis Control Branch,
  1984.

\bibitem{KappmeierGreen}
K.~Kappmeier~Green.
\newblock {\em Strategy for monitoring and sustainable integrated control or
  eradication of {G}lossina brevipalpis and G. austeni ({D}iptera: Glossinidae)
  in {S}outh {A}frica}.
\newblock PhD thesis, University of Pretoria, 2002.

\bibitem{KappmeierGreenPotgieterVreysen}
K.~Kappmeier~Green, F.~T. Potgieter and M.~Vreysen.
\newblock {\em Area--Wide Control of Insect Pests: From Research to Field
  Implementation}.
\newblock Editors: M. Vreysen, A. S. Robinson and J. Hendrichs. {S}pringer
  {D}ordrecht, {N}etherlands, 2007.

\bibitem{Torr1}
P.~M. Kgori, S.~Modo and S.~J. Torr.
\newblock The use of aerial spraying to eliminate tsetse from the {O}kavango
  {D}elta of {B}otswana.
\newblock {\em Acta Tropica}, 99:184--199, 2006.

\bibitem{MotloangMasumuVanDenBosscheMajiwaLatif}
M.~Y. Motloang, J.~Masumu, P.~Van Den~Bossche, P.~A.~O. Majiwa and A.~A.
  Latif.
\newblock Vector competance of field and colony {{\em {G}}}{\em lossina
  austeni} and {{\em {G}}}{\em lossina brevipalpis} for trypanosome species in
  {K}wa{Z}ulu--{N}atal.
\newblock {\em Journal of the South African Veterinary Association},
  80(2):126--140, 2009.

\bibitem{Nagel1}
P.~Nagel.
\newblock Environmental monitoring of tsetse control operations in {Z}ambia and
  {Z}imbabwe.
\newblock Technical report, Scientific Environmental Monitoring Group (SEMG)
  Project, 1993.

\bibitem{Parker1}
A.~Parker.
\newblock {\em By communication}.
\newblock 2008.

\bibitem{PerkinsAndRamberg}
J.~S. Perkins and L.~Ramberg.
\newblock Environmental recovery monitoring of tsetse fly sparaying impacts in
  the {O}kavango delta -- 2003.
\newblock Technical report, Harry Openheimer Okavango Research Centre,
  Botswana, 2004.

\bibitem{phelpsAndBurrows1}
R.~J. Phelps and P.~M. Burrows.
\newblock Puparial duration in {G}lossina morsitans orientalis under conditions
  of constant temperature.
\newblock {\em Entomologia Experimentalis et Applicata}, 12:33--43, 1969.

\bibitem{Randolph1}
S.~E. Randolph, D.~J. Rogers and F.~A.~S. Kuzoe.
\newblock Local variation in the population dynamics of {{\em {G}lossina
  palpalis palpalis}} ({R}obineau-{D}esvoidy) ({D}iptera: {G}lossinidae). ii.
  the effect of insecticidal spray programmes.
\newblock {\em Bulletin of Entomological Research}, 74:425--438, 1984.

\bibitem{RogersAndRandolph1}
David~J. Rogers and Sarah~E. Randolph.
\newblock Estimation of rates of predation on tsetse.
\newblock {\em Medical and Veterinary Entomology}, 4:195--204, 1990.

\bibitem{Turner1}
D.~A. Turner and R.~Brightwell.
\newblock An evaluation of a sequential aerial spraying operation against {{\em
  {G}lossina pallidipes}} {A}usten ({{\em {D}iptera: Glossinidae}}) in the
  {L}ambwe {V}alley of {K}enya: aspects of the post--spray recovery and
  evidence of natural population regulation.
\newblock {\em Bulletin of Entomological Research}, 76:331--349, 1986.

\bibitem{Vale3}
G.~A. Vale.
\newblock The responses of tsetse flies ({{\em {D}iptera: Glossinidae}}) to
  mobile and stationary baits.
\newblock {\em Bulletin of Entomological Research}, 64:545--588, 1974.

\bibitem{Williams1}
Brian~G. Williams, Robert~D. Dransfield and Robert Brightwell.
\newblock Tsetse fly ({D}iptera: Glossinidae) population dynamics and the
  estimation of mortality rates from life--table data.
\newblock {\em Bulletin of Entomological Research}, 80:479--485, 1990.

\end{thebibliography}

\end{document}